# Uncovering Sociological Effect Heterogeneity using Machine Learning

Jennie E. Brand* Jiahui Xu† Bernard Koch‡ Pablo Geraldo§

September 10, 2019

Abstract

Individuals do not respond uniformly to treatments, events, or interventions. Sociologists routinely partition samples into subgroups to explore how the effects of treatments vary by covariates like race, gender, and socioeconomic status. In so doing, analysts determine the key subpopulations based on theoretical priors. Data-driven discoveries are also routine, yet the analyses by which sociologists typically go about them are problematic and seldom move us beyond our expectations, and biases, to explore new meaningful subgroups. Emerging machine learning methods allow researchers to explore sources of variation that they may not have previously considered, or envisaged. In this paper, we use causal trees to recursively partition the sample and uncover sources of treatment effect heterogeneity. We use honest estimation, splitting the sample into a training sample to grow the tree and an estimation sample to estimate leaf-specific effects. Assessing a central topic in the social inequality literature, college effects on wages, we compare what we learn from conventional approaches for exploring variation in effects to causal trees. Given our use of observational data, we use leaf-specific matching and sensitivity analyses to address confounding and offer interpretations of effects based on observed and unobserved heterogeneity. We encourage researchers to follow similar practices in their work on variation in sociological effects.

* Professor of Sociology and Statistics, UCLA, Director, California Center for Population Research (CCPR), Co-Director, Center for Social Statistics, brand@soc.ucla.edu. ORCID: 0000-0002-6568-498X.

† Sociology Ph.D. Student, Pennsylvania State University, jiahuixu@ucla.edu. ORCID: 000-0003-2728-0674.

‡ Sociology Ph.D. Student, UCLA, bernardkoch@g.ucla.edu. ORCID: 000-0001-5312-3440.

§ Sociology Ph.D. Student, UCLA, pdgeraldo@ucla.edu. ORCID: 0000-0001-8352-9679.

[**] The authors benefited from facilities and resources provided by the California Center for Population Research at UCLA (CCPR), which receives core support (P2C-HD041022) from the Eunice Kennedy Shriver National Institute of Child Health and Human Development (NICHD). Brand presented versions of this paper at the American Sociological Association 2019 annual meeting, the International Sociological Association Research Committee on Social Stratification and Mobility (RC28) 2019 summer meeting, UCLA Women in Statistics Distinguished Lecture Series, Carolina Population Center Seminar Series at UNC-Chapel Hill, and Inequality and Social Policy Program at Harvard University. The ideas expressed herein are those of the authors.

# Author Biography Page

Jennie E. Brand is Professor of Sociology and of Statistics at UCLA. She is Director of the California Center for Population Research (CCPR) and Co-Director of the Center for Social Statistics at UCLA, and Chair of the Methodology Section of the American Sociological Association. She is also a member of the board of the International Sociological Association Research Committee on Social Stratification and Mobility, a member of the Board of Overseers of the General Social Survey, and a member of the Technical Review Committee for the National Longitudinal Surveys Program at the Bureau of Labor Statistics. Her research centers on social stratification and inequality and its implications for various outcomes that indicate life chances, with a methodological focus on causal inference.

Jiahui Xu is a Ph.D. student in Sociology at Pennsylvania State University. She earned her master's degree in Applied Economics at UCLA. Her research interests lie in causal inference and demography. She currently works on causal inference with machine learning and evaluation of sociological effect heterogeneity.

Bernard Koch is a Ph.D. student in Sociology at UCLA. He is interested in the science of science, cultural evolution, and computational methods. He is currently focused on the application of deep learning to network and causal inference problems to help identify how we can make science more equitable, efficient, and productive.

Pablo Geraldo is a Ph.D. student in Sociology at UCLA and student affiliate of the California Center for Population Research (CCPR). His research examines inequality in education and the labor market, using a mixture of causal inference, network analysis, and machine learning approaches.

# 1 Introduction

Population heterogeneity is pervasive. Individuals differ not only in pretreatment characteristics of interest to social scientists, (i.e., pretreatment heterogeneity), but also in how they respond to a common treatment, event or intervention (i.e., treatment effect heterogeneity). Treatment effect heterogeneity has important implications for social research and policy. The study of treatment effect heterogeneity can yield valuable insights into how scarce social resources are distributed in an unequal society (e.g., Brand 2010; Brand and Xie 2010; Heckman, Urzua, and Vytlacil 2006), how events differentially impact populations with different expectations of their occurrence (e.g., Brand and Simon Thomas 2014; Brand et al. 2019; Clark 2010; Turner 1995), and what may explain response heterogeneity, including differential selection (e.g., Heckman and Vytlacil 2007; Zhou and Xie [forthcoming]; Zhou and Xie 2019). If policymakers understand patterns of treatment effect heterogeneity, they can more effectively assign different treatments to balance competing objectives, such as reducing costs and maximizing average outcomes (Davis and Heller 2017).

Sociologists routinely partition their samples into subgroups by individual characteristics to explore how the effects of particular events or interventions vary across the population. Researchers often, for example, assume that effects vary by race and gender and indicators of socioeconomic status. Despite their ubiquity, such interactions may not represent the most meaningful variation in effects, or the partitions that are most consequential for a relationship of interest. Indeed, many researchers report stratified estimates by gender or race when the differences between the groups are not statistically or substantively significant. Longstanding theoretical priors, strong convention, and biases that we should examine differences by particular characteristics often drive these decisions. The practices used by researchers when examining heterogeneity via interaction effects also regularly fail to consider the causal assumptions and possible alternative interpretations based on selection underlying subpopulation differences in estimated effects (Kaufman 2018). Social scientists interested in causal inference also explore how effects vary by the likelihood of selection into treatment, by stratified analyses by propensity score strata, non-parametric methods, or exploring variation across different parameters of interest that indicate selection into treatment (Brand and Simon Thomas 2013; Heckman, Urzua, and Vytlacil 2006; Xie, Brand and Jann 2012). The interpretation of such analyses depends on both observed and unobserved selection into treatment (Brand et al. 2019; Brand and Xie 2010; Heckman and Vytlacil 2007; Zhou and Xie [forthcoming]; Zhou and Xie 2019). In both covariate and propensity partitioning methods, however, analysts determine the key subgroups.

Empirical manuscripts are largely written to suggest that decisions about which subgroups to explore occur before any data analyses. Indeed, much social scientific inquiry labors under the delusion that methods of discovery reflect inspiration from an unknown source, rather than from extensive trial and error analyses. Yet it is often difficult to know ex ante the subgroups most responsive to treatment. In social scientific practice, researchers routinely explore their data, running at times tens or hundreds of regressions to determine if subgroups of potential interest show meaningful differences in effect estimates. Researchers then proceed to report the effect estimates of those that do. If researchers select which interactions to report as a result of such analyses, and do not draw on cross-validation procedures or multiple-testing adjustments, they are subject to incorrectly rejecting a correct null hypothesis. That is, ad hoc searches for responsive subgroups may reflect noise within the data rather than true response variation. Moreover, covariates may be most informative when considered jointly, in complex and non-linear ways (e.g., upper-income whites with strong religious beliefs, rather than whites). It may be unclear which of the large number of possible joint covariates and covariate thresholds are best to consider before analyses.

We argue for an alternative data-driven approach that will inform researchers as to essential sources of effect heterogeneity and more transparently depict the analyses that led to a focus on particular groups. Machine-learning methods allow us to explore sources of variation that we may not have considered, i.e., to explore meaningful data-driven treatment effect heterogeneity. Causal trees, decision trees adapted to determine treatment effect heterogeneity (Athey and Imbens 2016), allow researchers to uncover subpopulations of interest that they had not prespecified with greater flexibility by searching over high dimensional functions of covariates. Using this approach, analysts can generate uncover key subpopulations that may or may not accord with conventional sociodemographic partitions and theoretical priors. Athey and Imbens (2016) also propose honest estimation, whereby they split the data into two subsamples for testing and estimation.

We demonstrate the application of causal trees to analyze a key topic in the social inequality literature, the distributional effects of college on low wage work over the life course. As we base our analyses on observational data, like much work of interest in sociology, we merge matching methods with causal trees, and use sensitivity analyses to explore the impact of unobserved confounding. We consider alternative interpretations of effects based on both observed and unobserved variables. Covariate and propensity-based partitioning methods offer a point of comparison to recursive partitioning based on causal trees. We demonstrate that typical interaction analyses would not have identified some of the most responsive subgroups. We encourage researchers to follow similar practices in their work on exploring variation in sociological effects, and offer straightforward guidelines by which to do so.

## 2 Heterogeneous Treatment Effects in Observational Data

We focus on heterogeneity in treatment effects based on observational data. Observational data can identify causal associations of social processes that are not easily subject to experimental manipulation (Lederer et al. 2019). Suppose we have units $i = 1, \dots n$, a pretreatment covariate vector $X_i$, a response $Y_i$, and treatment $D_i \in \{0, 1\}$. We assume potential outcomes for each unit ($Y_i^0$, $Y_i^1$) and define the unit-level treatment effect as:

$$\tau_i = {Y_i}^1 - {Y_i}^0, \tag{1}$$

where the observed outcome for unit $i$ is a potential outcome corresponding to the treatment received, and we never observe both outcomes. We define the average treatment effect ($ATE$) as:

$$ATE = E(\tau_i) = E({Y_i}^1 - {Y_i}^0). \tag{2}$$

We invoke an "unconfoundedness" or "selection on observables" assumption that conditional on $X$, there are no additional confounders between the treatment and the outcomes of interest. As it is generally infeasible to condition on $X$ in a fully nonparametric way, methods for estimating treatment effects under unconfoundedness often entail treating nearby units in the $x$-space as matches for the target treated unit.

One approach to determine nearby cases is to use the propensity score to approximate the assignment mechanism (Imbens and Rubin 2015). A propensity score is the probability of treatment conditional on a set of observed covariates:

$$e(x) = pr(D_i = 1 | X_i = x). \tag{3}$$

The score provides a summary measure of selection into treatment. We can use the propensity score to estimate a conditional average treatment effect ($CATE$) as:

$$CATE = \tau(x) = E\left[Y_i^{\,1} - Y_i^{\,0} \middle| e(x)\right].$$

(4)

If we know $e(x)$, we can obtain an unbiased estimate of $\tau(x)$ using methods like inverse $e(x)$ weighting or matching on $e(x)$. We also attend to common support, or overlap, across treated and control units, such that we eliminate treated/control units that have no comparable control/treated units. Using observational data, it is also important to relax the unconfoundedness assumption and conduct sensitivity analyses for hidden confounding (Rosenbaum 2002).

## 2.1 Heterogeneous *CATE*: Covariate and Propensity Partitioning

Our goal is to identify how treatment effects vary across the population. Common practice is to examine interactions or stratified analyses of the treatment indicator with selected theoretically-motivated covariates. Researchers routinely use regression interaction terms or covariate stratified analyses to explore subgroup variation. Yet assumptions necessary for causal inference are often unarticulated by sociologists exploring stratified models. That is, researchers using observational data may (implicitly) assume unconfoundedness for each subpopulation estimated treatment effect, without considering possible differential selection bias across groups. They also typically assume that there is common support between treated and control units within subgroups, without articulating or assessing the validity of that assumption. Another approach by which to assess treatment effect heterogeneity is to partition the sample into strata of the estimated propensity score and estimate propensity score-based *CATE*s to determine whether subpopulations with lower or higher estimated probabilities of treatment differ in their effect of treatment (Brand and Simon Thomas 2013; Xie, Brand, and Jann 2012). Researchers may explicitly partition the data to ensure adequate balance between covariate and propensity distributions in the treated and control subsamples, and ensure common support. In contrast to the approach of variation by $X$, partitioning by *e(x)* focuses our attention on potential differential selection bias in estimated subpopulation treatment effects.

## 2.2 Heterogeneous $CATE$: Recursive Partitioning Based on Machine Learning

We next assess treatment effect heterogeneity using a machine learning approach based on decision trees. Machine learning is a computational and statistical approach to extracting patterns and trends from data (Brand, Koch, and Xu [forthcoming]). Supervised learning algorithms learn from training data that contains both independent and dependent variables.[i] Some common supervised learning approaches include generalized linear models, kernel methods, artificial neural networks, and decision trees. Supervised learning algorithms can learn to predict outcomes. In learning the model, there is a tradeoff between a model's ability to achieve the optimal in-sample fit and its ability to generalize to new data. An overfit model fits too closely to the data, yielding a flexible model that accurately explains idiosyncratic patterns (i.e., noise) in those data and does not generalize well to new data. A learning algorithm must be flexible enough to fit the training data, while simultaneously not so complex that variance is high when fit to new data. Regularization strategies, like cross-validation, prevent overfitting and improve generalization. During training, supervised learning algorithms optimize in-sample performance for a loss function (also called objective or cost function), often the mean-squared error (*MSE*) for regression tasks. After training, researchers use evaluation metrics to assess out-of-sample predictive performance of the model.

*2.2.1. Causal Trees.* Statisticians and social and computer scientists have recently made progress in merging machine learning methods and causal inference (Athey 2019;

Brand, Koch, and Xu [forthcoming]). Scholars of causal inference are adopting approaches from machine learning while still attending to identification strategies from statistics and econometrics. Estimation of treatment effect heterogeneity using decision trees represents an especially promising use of machine learning methods for causal inference (Athey and Imbens 2017). Decision trees are a widely-used non-parametric machine learning approach that recursively partition the data by covariates into increasingly smaller subsets where data bear greater similarity (i.e., have a smaller variance or entropy or Gini coefficient) (Breiman et al. 1984). A tree represents the resulting hierarchical data structure. Regression trees are non-parametric approaches to fitting a model for the conditional expectation function through recursive, binary partitioning of data. A covariate and threshold are selected that minimize the in-sample loss function (e.g., the mean-squared error [*MSE*]) within the data, and the algorithm splits the sample into two new partitions. Cross-validation is used to construct a tree that maximizes predictive power without overfitting the data. Decision trees minimize the loss function only on the current subset of data at each split, rather than on the entire dataset. This "greedy" approach is computationally efficient, but may result in a locally optimal rather than globally optimal solution. Decision trees typically use adaptive estimation, relying on the same data for constructing partitions and predicting leaf-specific outcomes.

In standard decision trees, each leaf represents the average value of $Y$ for units in that leaf. If there are $k$ covariates and $N$ observations, we partition the covariate space $\mathbb{X}$ into $M$ mutually exclusive leaves $l_1$, … $l_M$ where we estimate the outcome for an individual with covariate $x$ in leaf $l_M$ as the mean of the outcome for training observations in that leaf. Let $x_j$ be a splitting covariate and $s$ be a split threshold, and let $l_1(j, s) = \{X \mid X_j \leq s\}$ and $l_2(j, s) = \{X \mid X_j > s\}$. The algorithm selects $\{j, s\}$ such that:

$$\min_{j,s}\left[\sum_{x_i\in l_1(j,s)}(Y_i-\bar{Y}_1(j,s))^2 \;+\sum_{x_i\in l_2(j,s)}(Y_i-\bar{Y}_2(j,s))^2\right], \tag{5}$$

where $\bar{Y}_1(j,s)$ and $\bar{Y}_2(j,s)$ are the mean outcomes in $l_1(j, s)$ and $l_2(j, s)$ respectively. This process is repeated recursively within each partition until a regularization penalty limits the depth of the tree. The resulting leaves contain a group of units with similar values of $Y$.

A new approach developed by Athey and Imbens (2016) extends decision trees to estimate causal effects. Applying a potential outcome approach to decision trees requires altering the objective. That is, in a causal tree, we want the best prediction of, not the outcome $Y$ as in the standard regression tree algorithm, but the *CATE*. The causal tree algorithm is thus an adaptation of regression trees for causal inference that attempts to partition the data in a way that minimizes heterogeneity within leaf-specific estimated average treatment effects, rather than minimizing heterogeneity within leaf-specific outcomes. The difficulty in predicting the leaf-specific treatment effect is that we have no "ground truth," or no observed value of the true treatment effect as we do when predicting the value of an observed outcome $Y$. This issue reflects the fundamental problem of causal inference (Holland 1986), i.e., that we do not observe the causal effect for any unit. To parallel the observed outcome approach, however, we can generate a treatment effect for each observation (Athey and Imbens 2016).

In addition to adapting the objective to maximize effect heterogeneity across leaves, Athey and Imbens (2016) advance "honest," rather than adaptive, estimation. In honest estimation, we do not use the same data for selecting the partitions of the covariate space and for estimation of leaf-specific effects. Once we construct a tree using a training sample $S^{tr}$, we estimate leaf-specific treatment effects using an estimation sample $S^{es}$.[ii] Notably, the criteria for constructing the partitions and cross-validation change in anticipation of honest estimation. Athey and Imbens (2016) introduce a modified expected *MSE* for the loss function that accounts for both honest estimation and the move to minimizing the *MSE* of treatment effects rather than outcomes:

$$\widehat{MSE}_{\tau(x)} = -\frac{1}{N^{tr}} \sum_{i \in S^{tr}} \hat{\tau}^2(X_i S^{tr} \Pi) + \left(\frac{1}{N^{tr}} + \frac{1}{N^{es}}\right) \sum_{l \in \Pi} \left( \frac{S^2_{S^{tr}(1)}(l)}{e(x)} + \frac{S^2_{S^{tr}(0)}(l)}{1 - e(x)} \right), \tag{6}$$

where $N^{tr}$ and $N^{es}$ is the sample size of the training sample and estimation sample, respectively, $\Pi$ is a potential partition of the covariate space[iii], and $S^2_{S^{tr}(1)}(l)$ and $S^2_{S^{tr}(0)}(l)$ are the sample variances for the treated and control units in leaf *l*, respectively. Honest estimation accounts for the uncertainty associated with the yet to be estimated leaf-specific treatment effects by including a penalty term for leaf-specific variance. For cross-validation, we use the same expression in the training sample. Honest estimation enables standard asymptotic properties in leaf-specific treatment effects.[iv]

While causal trees can find heterogeneous effects, they cannot however guarantee that confounding within leaves is addressed in observational studies. We must assume that leaf-specific estimates have sufficient overlap and are unbiased by differential selection into treatment. In observational studies, we should further adjust for covariate imbalance in estimated leaf-specific treatment effects (Athey and Imbens 2016). To do so, we adjust for covariates as we build the tree in $S^{tr}$ and when we estimate final leaf-specific effects in $S^{es}$ using nearest neighbor matching on the linearized propensity score (i.e., $logit(\hat{e}(x))$)[v] with four control units per treated unit. We estimate treatment effects in the causal tree by the difference in average outcomes between the treated and control observations within leaves:

$$CATE(\Pi) = \tau(x, \Pi) = E\left[Y_i^{\,1} - Y_i^{\,0} \middle| X_i \in l(x, \Pi)\right]. \tag{7}$$

We note that alternative approaches for adjustment are possible (Abadie and Imbens 2006; Imbens and Rubin 2015).

The causal tree algorithm proceeds as follows: (1) Draw a random subsample for training $S^{tr}$ and retain a holdout sample for estimation $S^{es}$; (2) Grow a tree via recursive partitioning in $S^{tr}$ that maximizes heterogeneity across leaves and minimizes heterogeneity within leaves, using cross-validation and matching within leaves; and (3) Estimate leaf-specific treatment effects in $S^{es}$ using matching within leaves. We depict this causal tree algorithm workflow in Figure 1.

*2.2.2. Causal Forests.* Compared to many other machine learning approaches, decision trees are an attractive tool for sociological applications because the criteria used to make predictions are transparent to the end-user. That is, the ability to plot the decision pathways of a regression tree render it a powerful tool not just for uncovering treatment effect heterogeneity, but also for interpreting and visualizing that heterogeneity. A disadvantage of single decision trees is that their greedy optimization produces unstable solutions. A reported tree may not be the only valid tree or the optimal tree. Different sample splits can result in different partitions and tree structures. An approach to avoid overfitting is to train multiple trees – an ensemble, or random forest – and average their predictions. In the random forest algorithm (Breiman et al. 1984; Breiman 2001), each tree in the forest is constructed by bootstrap aggregating (i.e., "bagging") the data by repeatedly resampling training data with replacement and generating a consensus prediction. Even with bagging, greedy trees tend to use the same features for similar decision sequences. Random forests thus combine bagging with a covariate resampling scheme that forces greedy trees to explore different decision sequences with other covariates. In other words, at each split, a given tree in the forest can only choose from a random subset of covariates.

Extending upon causal trees and random forests, Wager and Athey (2018) propose a causal forest for estimating treatment effects in the potential outcomes framework, assuming unconfoundedness.[vi] They use subsamples of the training data to grow the trees used to build the forest. The causal forest generates an ensemble of $B$ causal trees, each with an estimate of $\widehat{\tau_b}(x)$ and averages estimates from individual trees:

$$\hat{\tau} = B^{-1} \sum_{b=1}^{B} \widehat{\tau_b}(x) \,. \tag{8}$$

As the trees are not grown using data on the outcome, we have honest estimation without the use of sample splitting. Wager and Athey (2018) establish conditions under which predictions made by forests are asymptotically unbiased and normal. A random forest approach, however, does not give us a single, easily interpretable tree depicting treatment effect heterogeneity. We can, however, construct a metric of covariate importance by assessing the covariates chosen most often by the causal forest algorithm (i.e., a count of the proportion of splits on the variable of interest to a depth of four with a depth-specific weighting) and thus the strongest determinants of the structure of the trees in the forest (O'Neill and Weeks 2018). Additionally, as the causal forest estimator is similar to an adaptive nearest neighbor method where the trees generate the covariate space for selecting nearest neighbors, causal forests provide an alternative estimation strategy for estimating population *CATE*s. We use this approach in our estimation of the *CATE* alongside nearest neighbor matching.

## 2.3 Sensitivity Analyses for Heterogeneous Effects

In this paper, we rely on observational data and the unconfoundedness assumption to identify treatment effects. We assess how sensitive our estimated *CATE*s are to unobserved confounding by quantifying how the results obtained under the unconfoundedness assumption would change if we relaxed that assumption. To do so, we subtract a bias factor from the point estimate and confidence interval of the heterogeneous treatment effects obtained under unconfoundedness (Arah 2017; Gangl 2013; VanderWeele and Arah 2011). The bias term is equal to the product of two (partition-specific) parameters:

$$B = \gamma\lambda, \tag{9}$$

where

$$\gamma = E(Y|U = 1, D = d, e(x)) - E(Y|U = 0, D = d, e(x)) \tag{10}$$

and

$$\lambda = P(U = 1|D = 1, e(x)) - P(U = 0|D = 0, e(x)). \tag{11}$$

That is, $\gamma$ is the mean difference in the outcome associated with a unit change in an unobserved binary confounder, $U$, and $\lambda$ is the mean difference in the unobserved confounder between treated and control units.

## 3 Empirical Application

To demonstrate the approach, we assess heterogeneity in the effect of college on reducing low wage work over the career. The effects of college on wages is a key area of interest in social inequality research (Hout 2012). By focusing on low wage work, we shift attention to how college may circumvent disadvantaged labor market outcomes for particular subpopulations. There is a great deal of rhetoric about limiting college for segments of the population, particularly more disadvantaged students on the margin of school continuation. If we observe benefits for this population that may match, or even exceed, those of more traditional college students, we gain insight into whether college pays off for this subpopulation of potential college-goers. We draw on observational data and a highly selective treatment condition, completing college, in order to illustrate the use of causal trees with non-experimental data. We address four research questions: (1) Does college reduce the proportion of time in low wage work over the career? (2) Does the effect of college on low wage work vary by key covariates that influence the likelihood of completing college (i.e., propensity of college, parental income, mothers' education, measured ability, and race)? (3) Does the effect of college on low wage work vary by other covariates or combinations of covariates that we did not specify? That is, do causal trees uncover subpopulations that differ in their effects of college on wages? (4) How sensitive are the subpopulation treatment effect estimates to unobserved confounding?

## 3.1 Data and Descriptive Statistics

We use data from the Bureau of Labor Statistics 1979-2014 waves of the National Longitudinal Survey of Youth (NLSY) 1979 cohort. This nationally representative longitudinal data provide information on respondents' sociodemographic background, achievement, skills, educational attainment, and long-term earnings trajectories from youth to late-career, and has been used widely in assessing the effects of college on wages. We restrict the sample to individuals who were 14-17 years old at the baseline survey in 1979 ($n$ = 5,582) and who had completed at least the 12th grade ($n$ = 4,548). These sample restrictions ensure that all variables we use to predict college are measured pre-college, and that we compare college completers to those who completed at least a high school education. About one-fifth of the sample completed college by age 25. We focus on the proportion of time spent in a low wage job for the years from 1990 to 2014, when respondents were roughly between the ages of 25 and 50. We measure low wage work as less than two-thirds of the median hourly wage for that year (Presser and Ward 2011). In Table 1, we report covariate means by college completion. Descriptive statistics on our pre-college covariates suggest well-documented socioeconomic differences in educational attainment.[vii]

– TABLE 1 ABOUT HERE –

## 3.2 Estimating Propensity Scores

To estimate the propensity of college, we adopt an iterative procedure suggested by Imbens and Rubin (2015). We begin with a set of baseline covariates $K_B$, which includes theoretically key predictors of college which also plausibly relate to wages. These include covariates used in several papers of college effects by Heckman and colleagues (e.g., Carneiro, Heckman, and Vytlacil 2011): race, parents' education and income, family structure, residence, and cognitive ability.[viii] In the second step, we consider $K - K_B$ additional possible covariates in turn, where we add the covariate with the largest likelihood ratio statistic that exceeds a pre-set constant $C_L$ (1.0, or a *z*-statistic of 1.0), and the process repeats. We consider additional measures of family background, psychosocial skills, school characteristics, and family formation. This step involved 176 logistic regressions, and a resulting model with 22 linear terms.[ix] In the third step, we assess which of all possible higher order and interaction terms $[K_L\ (K_L + 1)/2]$ (i.e., 253 possible additional terms) to include in the model. We then follow the same procedure as above, where we add the term $K_Q$ with the largest likelihood ratio statistic that exceeds the pre-set constant $C_Q$ (*z*-statistic of 1.96). This procedure involved 3,527 regressions. The resulting model includes 22 linear terms, one higher-order term, and 12 interaction terms. We estimate the propensity score with the vector of $1 + K_L + K_Q$ terms using a logit regression model; results are reported in Appendix Table A.

We eliminate units outside the region of common support, i.e. treated cases with values higher than the highest propensity score among the controls ($\hat{e}(x) = 0.923$) and values lower than the lowest propensity score among the treated ($\hat{e}(x) = 0.003$) ($n = 4{,}085$). The propensity score we estimate here has a high correlation with covariate balancing and machine learning methods: We observe a correlation of 0.97 with a propensity score generated by a covariate balancing method proposed by Imai and Ratkovic (2014) and a correlation 0.94 with a propensity score generated by a random forest.[x]

## 3.3 *CATE* Estimate

In Table 2, we provide the full population *CATE* estimate using nearest neighbor matching based on the linear propensity score with four control units per treated unit, and using causal forests. We find that college completion is associated with a significant 18 percentage point reduction in the proportion of time spent in a low wage job across the career based on matching, and a 17 percentage point reduction based on causal forests. Figure 2 is a plot of the covariate balance we achieve for matching and causal forests, and For most covariates, both matching and causal forests eliminate mean differences.

– TABLE 2, FIGURE 2 ABOUT HERE –

Figure 3 is a histogram of *CATE* estimates from matching, which suggests a distribution of treatment effects. We can also construct individual treatment effects (*ITE*s) using nearest neighbor matching methods and causal forests. In Figure 4, we present a distribution of *ITE*s from our causal forest with 95 percent confidence intervals. The dashed red lines indicate quartiles of the *ITE* distribution. The largest effects in the first quartile fall around -0.25, with effects in the fourth quartile around -0.10. We next proceed to identify subpopulations for which estimates differ in the distribution represented in Figures 4 and 5.

– FIGURES 3, 4 ABOUT HERE –

## 3.4 Heterogeneous *CATE* Estimates: Covariate and Propensity Partitioning

We examine stratified effects of college completion by several *a priori* theoretically-motivated covariates: parental income, mothers' education, ability, and race. These covariates strongly influence selection into college and indicate levels of socioeconomic advantage. We divide parental income and ability into terciles of the distributions, divide mothers' education into categories of less than high school, high school degree, and some college or more, and divide respondents' race into black, Hispanic, and white. We also construct three propensity score strata to assess effects for the low, middle, and high propensity college goers, where low ranges from 0 to less than 0.2, middle from 0.2 to less than 0.5, and high from 0.5 to 1. We report estimated effects via stratified models based on matching in Table 3. We find larger effects of college on reducing low wage work for those with a low propensity to complete college, low ability, low income, low mothers' education, and blacks. The effects of college on low-wage work for the most advantaged individuals appear close to zero. We report tests of significance between estimated coefficients in Appendix Table B. Most estimates significantly differ from one another.

– TABLE 3 ABOUT HERE –

In these analyses, we invoked the unconfoundedness assumption. Whether this is a reasonable assumption is a substantive issue, which depends upon the quality of the covariates in capturing potential selection bias. We recognize that even with a rich set of pretreatment covariates, potential confounders remain (e.g., motivation or idleness). In Table 4, we report sensitivity bounds on the estimated coefficients reported in Table 3. The effect reaches non-significance when the unobserved confounder has a sizable difference between individuals who do and do not complete college ($\lambda$) or a strong effect on low wage work ($\gamma$). Suppose, for example, that idleness, unobserved in our data, increases the proportion of time in low wage work over the career, and is lower among individuals who complete college than those who do not. When $\lambda$ equals -10 percent, we assume that the prevalence of idle individuals is 10 percent lower in the college-educated group than in the non-college-educated group. When $\gamma$ equals 10 percent, we assume that idle individuals have a 10 percent higher level of being in low wage work relative to those who are not idle (all else equal).[xi] We let the values of $\gamma$ range from 10 to 40 percent, and fix the value of $\lambda$ at -10 percent.[xii] We observe that the effect of college on reducing low wage work remains significant for the most disadvantaged college completers at each value we consider, even when unobserved differences have a substantial impact on low wage work ($\gamma = 40$) and the prevalence of the unobserved factor differs between college and non-college graduates by 10 percent ($\lambda = -10$). Estimates also remain significant for the middle propensity score, parental income, and middle and high mothers' education subpopulations, and for Hispanics and whites. Effects among individuals with a high propensity of college and high parental income are more sensitive to confounding.

– TABLE 4 ABOUT HERE –

In Table 5, we also offer an alternative stratified approach based on causal forests in which we present means of covariates for each quartile of the causal forest *ITE* estimates represented in Figure 5 (O'Neill and Weeks 2018; Seibold, Zeileis, and Hothorn 2019). Results suggest that those on the margin of college completion have higher representation in the first quartile of treatment effects, i.e., the most responsive individuals. We find that fewer than 1 percent of individuals have a mid- or high propensity in the first quartile while 24 percent of individuals have a low propensity of college (i.e., over 97 percent of those cases in the first quartile). Patterns are similar, yet not as strong, for parental income, ability, and mothers' education. That is, the most responsive individuals are those with low ability, levels of mothers' education, and parental income. Blacks have a high proportion in the first quartile, while whites evenly distribute across quartiles.

– TABLE 5 ABOUT HERE –

### 3.5 Heterogeneous *CATE* Estimates: Recursive Partitioning

Figure 5 depicts the results of the causal tree for the impact of college completion on the proportion of time in low wage work (estimates are also reported in Appendix Table C). We include the 22 covariates described in Table 1 as well as the estimated propensity score as input splitting covariates. We limit the depth of the tree by requiring 50 treated and control units per leaf.[xiii] We use 50 percent of the sample to train the data and grow the tree structure, and reserve the remaining 50 percent of the sample as a holdout sample for estimation of leaf-specific treatment effects within that tree. The causal tree is color-coded to indicate the size of the association, with blue representing larger negative effects and yellow smaller effects (nearing zero). We examine both partitions that result in clustering of branches with leaves that are blue versus yellow, and particular leaves with large effects. We use matching to estimate leaf-specific treatment effects.[xiv]

The primary division depicted in Figure 5 occurs for mothers' education, with individuals whose mothers had less than a high school degree having larger negative effects of college on time spent in low wage work. Individuals whose mothers have less than a high school degree have a 26 percentage point reduction in low wage work, yet the largest effect accrues to those whose mothers do not complete high school and who grew up in large families – a 33 percentage point reduction in low wage work. Individuals whose mothers had at least a high school degree have smaller effects of college on low wage work. The yellow branches to the right demonstrate this pattern. Notably, individuals with mothers with at least a high school degree and higher family income (greater than $15,000 per year in the late 1970s) have the smallest effects, with the largest effect on this branch among those who do not expect to attend college (about 14 percentage point reduction in low wage

work). For individuals with mothers with at least a high school degree and lower family income, we observe a relatively large effect (a 19 percentage point reduction), but particularly among those whose fathers do not attend college (21 percentage point reduction). Appendix Table D provides tests of significance across leaves, suggesting significant differences across most leaves.

– FIGURE 5 ABOUT HERE –

In interpreting the results of the causal tree, we focus less on the the selected covariates that we partition the sample by, and more on the characteristics of the subpopulations that we determine are most responsive to treatment. Consider the most responsive leaf (leaf 2), those whose mothers did not complete high school and who grew up in large families. Almost half of the sample is black, and another third is Hispanic. About 40 percent did not grow up with both parents, and almost two-thirds had engaged in some delinquent activity. About 60 percent grew up in the south. These individuals have a relatively low propensity for college. Table 6 provides leaf-specific propensity score descriptive statistics and balance metrics for all of the leaves in the causal tree. We find that those subpopulations with a larger impact of college on reducing low wage work tend to be those with the lowest propensity for college, as we would expect given the propensity partitioned results. They also have higher propensity score imbalance; that imbalance is substantially, although not entirely, reduced via matching. Table 7 provides sensitivity bounds on the estimated effects. The leaf-specific estimates, particularly the sizable estimates associated with the most disadvantaged subpopulations, are robust to unobserved confounding. For example, the largest estimate in leaf 2 remains substantial even if the confounding variable reduced low wage work by 40 percent ($\gamma$) and differed by 10 percent among college and non-college graduates ($\lambda$).

– TABLES 6, 7 ABOUT HERE –

More disadvantaged subpopulations, those on the margin of school continuation, have larger effects of college on reducing low wage work. We identify this pattern across the various partitioning strategies. Yet the groups identified by the causal trees are not necessarily those we would identify by our theoretical priors. For example, while we consider strata based on mothers' education in Table 3, we did not consider individuals with mothers without a high school degree and who grew up in large families, nor those with high school educated mothers yet low parental income and fathers who did not attend college. The causal trees did not identify many dichotomous covariates, like race, as indicating key subpopulations, as the tree prefers to split on continuous covariates. We note, however, that the subpopulations identified have strong correlations with variables like race. This tree also did not identify the propensity score as a key partition, yet these subpopulations are highly correlated with those stratified by propensity scores.

As we note above, tree stability is a concern, i.e., we may get different trees if we generate different random splits of the training and test data. To test tree structure stability, we generate 100 causal trees with different random splits of the training and test data. We find that 89 percent of the time we get the exact tree structure we present above. We get six additional trees for the remaining 11 percent of trees whose structure differs by the depth of the tree. We also run causal forests with 4,000 trees. Each honest tree is fit using 50 percent of the data for generating partitions and 50 percent for estimating treatment effects. For each tree in the forest, we use a random subsample of the set of covariates for potential partitioning variables. Figure 6 is a plot of covariate importance, i.e., the strongest determinants of generating the structure of the trees in the forest. The $x$-axis indicates relative importance scores; we are concerned only with the relative strength across covariates. The results suggest that the estimated propensity score is most important, followed by parents' income, ability, fathers' education, school disadvantage, mothers' education, family size, social control, and college preparatory program. The remaining variables have minimal importance in terms of determining the structure of the trees. The covariates that generate the primary splits in the causal tree we present in Figure 6 are a subset of those identified here.

– FIGURE 6 ABOUT HERE –

## 3.6 Interpretation of Heterogeneous $CATE$ Estimates

We note above the possibility that the unconfoundedness assumption does not hold in observational data (Zhou and Xie [forthcoming], Zhou and Xie 2019). Continuing schooling is a highly selective process. Of the possible unobserved factors, some are systematic, reflecting individuals' resistance to continuing their schooling. Although we use sensitivity analyses above to address the possibility of unobserved confounding, let us consider here how unobserved confounding may influence the pattern of observed results. Let us denote unobserved resistance to college as $u$, as a summary measure of unobserved confounding. We describe the latent education function $D^*(\cdot)$ as:

$$D^* = e(x) - u. \tag{12}$$

Individuals complete college when $D^*(\cdot)$ exceeds 0:

$$D = \begin{cases} 1 & if\ D^* \geq 0, \\ 0 & otherwise. \end{cases} \tag{13}$$

The effect of college also varies by both $e(x)$ and $u$. In estimating heterogeneous treatment effects under unconfoundedness, we assume that the treatment effect varies by $e(x)$ and not by $u$. However, we can incorporate unobserved response heterogeneity in our interpretation of the effects we observe.

Let us define the effect of college on wages to be a function of both *e(x)* and $u$, i.e., the marginal treatment effect (*MTE*):

$$MTE = E[Y_i^1 - Y_i^0 | e(x), u]. \quad (14)$$

In Figure 7 (adapted from Zhou and Xie [forthcoming] and Brand et al. [2019]), we depict alternative ways in which we can interpret treatment effect heterogeneity. The darker shaded regions indicate a larger treatment effect magnitude (i.e., larger negative effects of college on proportion of time in low wage work). In Figure 7(a) and 7(b), we assume unconfoundedness but allow for effect heterogeneity by *e(x)*, i.e., college effects are assumed to vary by the propensity but not by unobserved resistance. In Figures 7(c) and 7(d), we consider the general case of equation (14) and allow *MTE* to be a function of both *e(x)* and *u*. Figures 7(a) and 7(c) depict effects for all units; Figure 7(b) and 7(d) depict effects for treated units, i.e., the subpopulation for which $e(x) > u$. We note the high correlation between *e(x)* and *u* among treated units in Figure 7(d): At low values of *e(x),* the estimated effect includes proportionally more individuals who have low values of resistance *u*; at high values of *e(x)*, the estimated effect includes more variation with respect to *u*, and thus proportionally more individuals who have high values of *u*. Thus, we can also interpret patterns of effects that we observe at low values of *e(x)* as applying to a particular subpopulation with low values of *u*. This interpretation corresponds to the propensity partitioning results, but also to the covariate and recursive partitioning results that bear high correlations with the propensity of college.

– FIGURE 7 ABOUT HERE –

# 4 Discussion

Heterogeneity in response to an event or intervention is common. We cannot reasonably presume that individuals respond identically to life events. We aim to understand heterogeneity, both in the characteristics that predispose some groups to experience particular events and how those characteristics govern differential response to those events. A critical question is whether social scientists know *a priori* which characteristics of individuals shape the distribution of responses. One long-standing approach in sociology is to determine subgroups of interest who we theorize should respond differently, and test those possibilities in our data. There are many advantages to doing so, as we may have theoretical interest in whether blacks or whites or men or women or those who grew up in disadvantaged versus more advantaged families are differentially affected by particular events. For example, we may want to know whether disadvantaged students benefit more or less from college than more advantaged students, because our policies target the recruitment of such students and we want to estimate the expected gain. We may likewise want to know whether students with a low propensity to complete college benefit more or less, as such knowledge of the stratification process informs us as to the consequences of the unequal distribution of scarce resources. Such analyses also give us insight into how selection into treatments may confound the relationships we observe across subgroups.

Yet often our data can tell us something that we had not thought of before analyses. Indeed, a great deal of the excitement of empirical social scientific work lies in unexpected discovery. Data-driven discoveries are common, yet the analyses by which sociologists typically go about them are problematic, and may not move us beyond our expectations, and inherent biases, to explore new meaningful groups. Most sociological analyses that explore covariate interactions neglect how combinations of covariates and nonlinear interactions may best identify key subpopulations of interest. In this paper, we use causal trees to uncover sources of treatment effect heterogeneity. We use honest estimation by using different subsamples for determining the model and estimating effects. Strategies like these will increasingly be needed to justify analytic decisions in applied work (Athey 2019). While Athey and Imbens (2016) developed causal trees using experimental data, we apply these methods to observational data, and use matching to address confounding. We compare results based on causal trees to traditional strategies based on covariate and propensity partitioning. Our empirical application addresses a central question in research on social inequality, the impact of college on wages. We identify sources of heterogeneity in effects, and unanticipated subgroups of notable interest. We use sensitivity analyses to address the possibility of lingering unobserved confounding, and offer interpretations of effects based on both observed and unobserved heterogeneity.

Our predetermined ideas as to which groups matter surely stifle social scientific progress. In this paper, we adopt a machine-learning based approach to studying causal effects that allows for the discovery of meaningful treatment effect heterogeneity, and avoids data-driven dangers commonly employed in practice. Machine learning algorithms are attractive for generating models where there may be numerous interaction effects *a priori* unknown to researchers. We urge sociologists interested in variation in effects to apply these techniques in order to engage more explicitly with methods of discovery and improve research practices for exploring effect heterogeneity.

# References

Abadie, Alberto and Guido Imbens. 2006. "Large Sample Properties of Matching Estimators for Average Treatment Effects." *Econometrica*, 74(1)235–267.

Arah, Onyebuchi. 2017. “Bias Analysis for Uncontrolled Confounding in the Health Science.” *Annual Review of Public Health* 38:12.1-12.16.

Athey, Susan. 2019. “The Impact of Machine Learning on Economics.” P. 507-547 in *The Economics of Artificial Intelligence: An Agenda*, Joshua Gans and Avi Goldfarb eds. Chicago: University of Chicago Press.

Athey, Susan and Guido Imbens. 2016. “Recursive Partitioning for Heterogeneous Causal Effects.” *Proceedings of the National Academy of Sciences* 113(27):7353-7360.

Athey, Susan and Guido Imbens. 2017. “The State of Applied Econometrics: Causality and Policy Evaluation.” *Journal of Economic Perspectives* 31(2):3-32.

Athey, Susan, J. Tibshirani, and Stefan Wager. 2019. “Generalized Random Forests.” *The Annals of Statistics* 47(2):1148-1178.

Brand, Jennie E. 2010. “Civic Returns to Higher Education: A Note on Heterogeneous Effects.” *Social Forces* 89(2):417-434.

Brand, Jennie E., Bernard Koch, and Jiahui Xu. [forthcoming]. “Machine Learning.” *SAGE Research Foundations*.

Brand, Jennie E., Ravaris Moore, Xi Song, and Yu Xie. 2019. “Parental Divorce is Not Uniformly Disruptive to Children’s Educational Attainment.” *Proceedings of the National Academy of Sciences* 116(15):7266-7271.

Brand, Jennie E. and Juli Simon Thomas. 2013. “Causal Effect Heterogeneity.” Pp. 189-214 in *Handbook of Causal Analysis for Social Research*, Stephen L. Morgan ed., Springer Series.

Brand, Jennie E. and Juli Simon Thomas. 2014. “Job Displacement Among Single Mothers: Effects on Children’s Outcomes in Young Adulthood.” *American Journal of Sociology* 119(4):955-1001.

Brand, Jennie E. and Yu Xie. 2010. “Who Benefits Most from College? Evidence for Negative Selection in Heterogeneous Economic Returns to Higher Education.” *American Sociological Review* 75(2):273-302.

Breiman, Leo, J. H. Freidman, R. A. Olshen, and C. J. Stone. 1984. *Classification and Regression Tress.* Monterey: Wadsworth & Brooks Cole.

Breiman, Leo. 2001. “Random Forests.” *Machine Learning* 45(1):5-32.

Carneiro, Pedro, James J. Heckman, and Edward Vytlacil. 2011. “Estimating Marginal Returns to Education.” *American Economic Review* 101(6):2754-2781.

Clark, A. 2010. “Boon or Bane? Others’ Unemployment, Well-being and Job Insecurity.” *Labor Economics* 17:52–61.

Davis, Jonathon M. V. and Sara B. Heller. 2017. “Using Causal Forests to Predict Treatment Heterogeneity: An Application to Summer Jobs.” *American Economic Review: Papers and Proceedings* 107(5):546-550.

Gangl, Markus. 2013. “Partial Identification and Sensitivity Analysis.” Pp. 377-402 in *Handbook of Causal Analysis for Social Research*, edited by S.L. Morgan. New York: Springer.

Grimmer, Justin, Solomon Messing, and Sean J. Westwood. 2017. “Estimating Heterogeneous Treatment Effects and the Effects of Heterogeneous Treatments with Ensemble Methods.” Working paper, April 14.

Hahn, P. Richard, Jared Murray, and Carlos Carvalho. 2017. “Bayesian Regression Tree Models for Causal Inference: Regularization, Confounding, and Heterogeneous Effects. *arXiv preprint arXiv:1706.09523.*

Hainmueller, Jens, Jonathon Mummolo, and Yiqing Xu. 2018. “How Much Should We Trust Estimates from Multiplicative Interaction Models? Simple Tools to Improve Empirical Practice.” *Political Analysis* 1-30.

Heckman, James, Sergio Urzua, and Edward Vytlacil. 2006. ‘‘Understanding Instrumental Variables in Models with Essential Heterogeneity.’’ *The Review of Economics and Statistics* 88:389–432.

Heckman, James, and Edward Vytlacil. 2007. “Econometric Evaluation of Social Programs, Part II: Using the Marginal Treatment Effect to Organize Alternative Econometric Estimators to Evaluate Social Programs, and to Forecast their Effects in New Environments." in *Handbook of Econometrics*, vol. 6, edited by J. Heckman and E. Leamer, chap. 71. Elsevier.

Hout, Michael. 2012. "Social and Economic Returns to College over the Life Course." *Annual Review of Sociology* 38:10.1-10.22.

Imai, Kosuke and Marc Ratkovic. 2014. "Covariate Balancing Propensity Score." *Journal of the Royal Statistical Society* 76(1):243-263.

Imbens, Guido and Donald Rubin. 2015. *Causal Inference for Statistics, Social, and Biomedical Sciences.* Cambridge: Cambridge University Press.

Lederer, David J. et al. 2019. "Control of Confounding and Reporting of Results in Causal Inference Studies: Guidelines for Authors from Editors of Respiratory, Sleep, and Critical Care Journals." *AnnalsATS* 16(1): 22-28.

Lee, Brian K., Justin Lessler, and Elizabeth A. Stuart. 2009. "Improving Propensity Score Weighting using Machine Learning." *Statistics in Medicine* 29:337-346.

McCaffrey, Daniel F., Greg Ridgeway, and Andrew R. Morral. 2004. "Propensity Score Estimation with Boosted Regression for Evaluating Causal Effects in Observational Studies." *Psychological Methods* 9(4):403.

Powers, Scott et al. 2018. "Some Methods for Heterogeneous Treatment Effect Estimation in High Dimensions." *Statistics in Medicine* 37:1767-1787.

Presser, Harriet and Brian W. Ward. 2011. "Nonstandard Work Schedules over the Life Course: A first Look." *Monthly Labor Review* July:3-16.

Nie, X. and Stefan Wager. 2017. "Quasi-Oracle Estimation of Heterogeneous Treatment Effects." Working paper *arXiv:1712.04912.*

O'Neill, Eoghan and Melvyn Weeks. 2018. "Causal Tree Estimation of Heterogeneous Household Response to Time-of-Use Electricity Pricing Schemes." Working paper *arXiv:1810.09179v1.*

Rosenbaum, Paul R. 2002. *Observational Studies.* Springer.

Seibold, Heidi, Achim Zeileis, and Torseten Hothorn. 2019. "model4you: An R Package for Personalized Treatment Effect Estimation." *Journal of Open Research Software* 17: DOI: https://doi.org/10.5334/jors.219.

Turner J. B. 1995. "Economic Context and the Health Effects of Unemployment." *Journal of Health & Social Behavior* 36:213–29.

VanderWeele, Tyler J. and Onyebuchi A. Arah. 2011. "Unmeasured Confounding for General Outcomes, Treatments, and Confounders: Bias Formulas for Sensitivity Analysis." *Epidemiology* 22(1):42-52.

Wager, Stefan and Susan Athey. 2018. "Estimation and Inference of Heterogeneous Treatment Effects using Random Forests." *Journal of the American Statistical Association* 113(523):1228-1242.

Westrish, Daniel, Justin Lessler, and Michele J. Funk. 2010. "Propensity Score Estimation: Neural Networks, Support Vector Machines, Decision Trees (CART), and Meta-Classifiers as Alternative to Logistic Regression." *Journal of Clinical Epidemiology* 63(8): 826-833.

Xie, Yu, Jennie E. Brand, and Ben Jann. 2012. "Estimating Heterogeneous Treatment Effects with Observational Data." *Sociological Methodology* 42:314-347.

Xie, Yu. 2007. "Otis Dudley Duncan's Legacy: The Demographic Approach to Quantitative Reasoning in Social Science." *Research in Social Stratification and Mobility* 25: 141-156.

Zhou, Xiang, and Yu Xie. [forthcoming]. "Marginal Treatment Effects from a Propensity Score Perspective." *Journal of Political Economy.*

Zhou, Xiang, and Yu Xie. 2019. "Heterogeneous Treatment Effects in the Presence of Self-Selection: A Propensity Score Perspective." *Sociological Methodology* 1-36. DOI: 10.1177/0081175019862593

[i] Unsupervised algorithms do not use data on dependent variables. Supervised learning tasks involving a continuous outcome are regression tasks, and those involving a categorical outcome are classification tasks.

[ii] Using adaptive estimation, spurious extreme values of the outcome (or in our case, the treatment effect) are likely to be placed into the same leaf as other extreme values, and thus the leaf-specific means or effects are more extreme than they would be in an independent sample (Athey and Imbens 2016). Loss of precision due to smaller sample size for estimation is overshadowed by the gain in minimizing bias.

[iii] Hahn et al. (2018) emphasize using the propensity score as one of the input covariates, and we follow their advice here.

[iv] Traditional decision trees are unconcerned with standard errors on leaf-specific treatment effects because interpreting leaf-specific effects is not the motivation of the construction of the tree.

[v] The linear propensity score is preferable to the raw score because the former does not penalize differences in pretreatment covariates at the tails of the propensity score distribution (Imbens and Rubin 2015). For example, on the raw propensity score scale, a treated unit with $\hat{e}(x) = 0.10$ is considered as close to a control unit with $\hat{e}(x) = 0.11$ as to a control unit with $\hat{e}(x) = 0.09$. But in terms of the covariates, the treated unit tends to be closer to the former than to the latter. The linear propensity score, by transforming $\hat{e}(x)$ back to the scale of the covariates, does not suffer from this issue.

[vi] Two procedures can be used to build causal forests: double sample trees and propensity trees (Wager and Athey 2018). In the first approach, we grow a tree using the $S^{tr}$ sample and then estimate $\hat{\tau}(x)$ on the $S^{te}$ sample. In the second approach, the outcome is the treatment assignment $D_i$, and leaves minimize heterogeneity in assignment to treatment, and we estimate $\hat{\tau}(x)$ in the propensity-partitioned leaves. Propensity trees bear similarity to the propensity partitioning approach (Xie, Brand, and Jann 2012).

[vii] Those who completed college are more likely to come from families with highly educated parents, high incomes, both parents present, and fewer siblings. They also have higher average cognitive test scores and are more likely to have enrolled in college-preparatory classes. They attend more advantaged high schools, have higher educational expectations and aspirations, and have friends with higher educational expectations. College graduates are also less likely to have started families during adolescence.

[viii] Ability is measured by the 1980 Armed Services Vocational Aptitude Battery (ASVAB), adjusted for age and standardized (Cawley et al. 1997). We also include a measure indicating whether data were imputed.

[ix] Knowledge of father's education, religion, school racial composition, self-esteem, and family attitudes did not reach the threshold for model improvement.

[x] Machine learning methods have been used to estimate $e(x)$ using CART, neural networks, and random forests (Lee, Lessler, and Stuart 2009; McCaffrey, Ridgeway, and Morral 2004; Westrich, Lessler, and Funk. 2010).

[xi] With the exception of expectations, aspirations, and college track, most of the pre-college indicators differ between college and non-college graduates by less than 5 percent. We would not expect many unobserved factors to differ between college and non-college graduates by more than 10 percent.

[xii] The sensitivity results when $\gamma$ is negative and $\lambda$ is positive are the same as those we present here, so there is no loss of information by not including the opposite sign.

[xiii] Larger leaves render results more consistent across samples, yet depict less heterogeneity.

[xiv] We offer an interactive online data visualization of the tree at https://causal-tree-svelte.tonyhschu.now.sh, The website, including the URL, does not identify the authors. An R Markdown file is available upon request, and we are developing Stata programs to implement these methods.

Figure 1. Causal Tree Algorithm Workflow

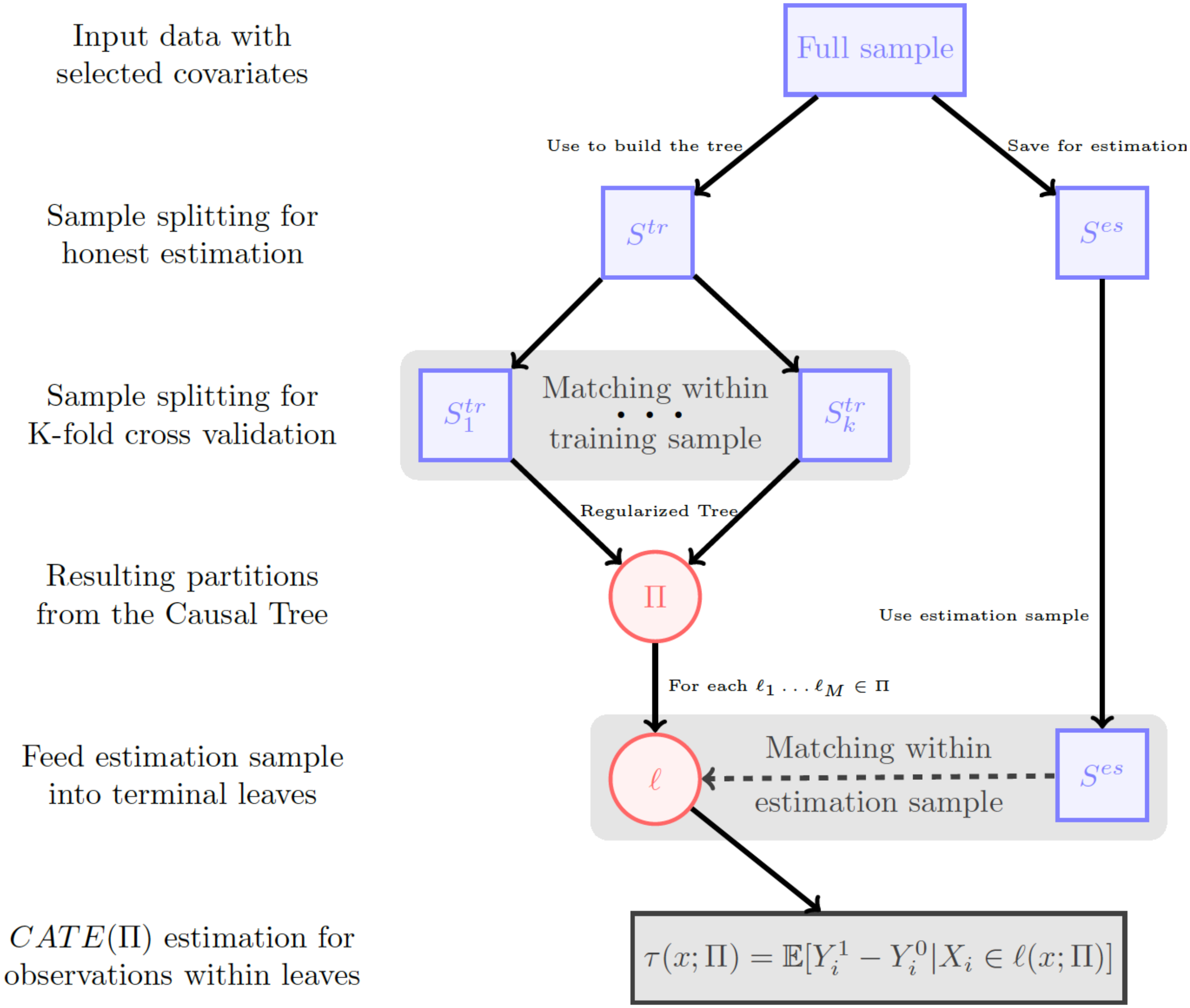

Figure 2. Covariate balance for matching and causal forests of the Effect of College Completion

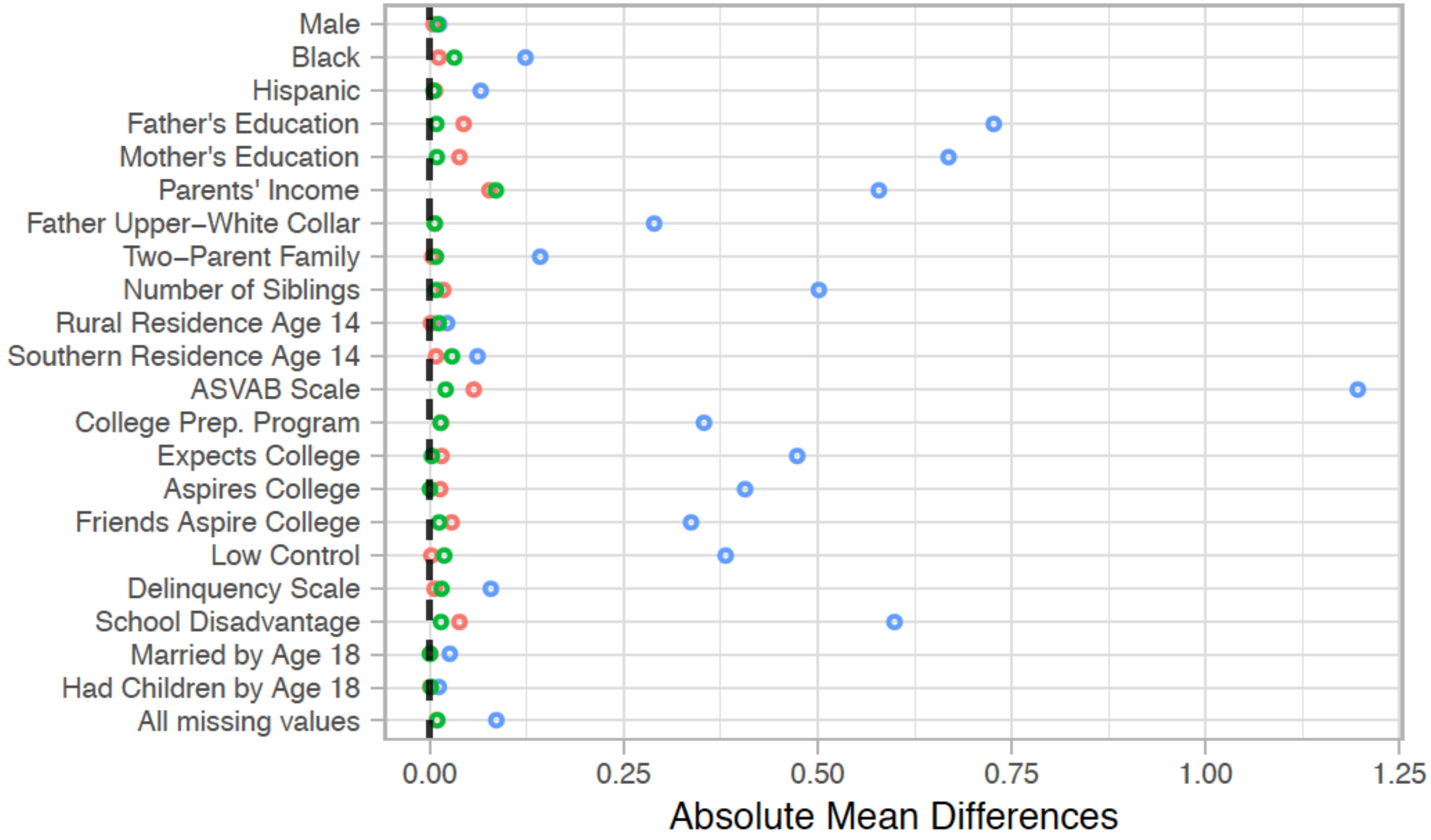

*Notes:* Blue: unadjusted; Green: Adjusted by matching; Red: Adjusted by causal forest;

Figure 3. Histogram of *CATE* estimates of the Effect of college completion on the proportion of time in low wage work from matching

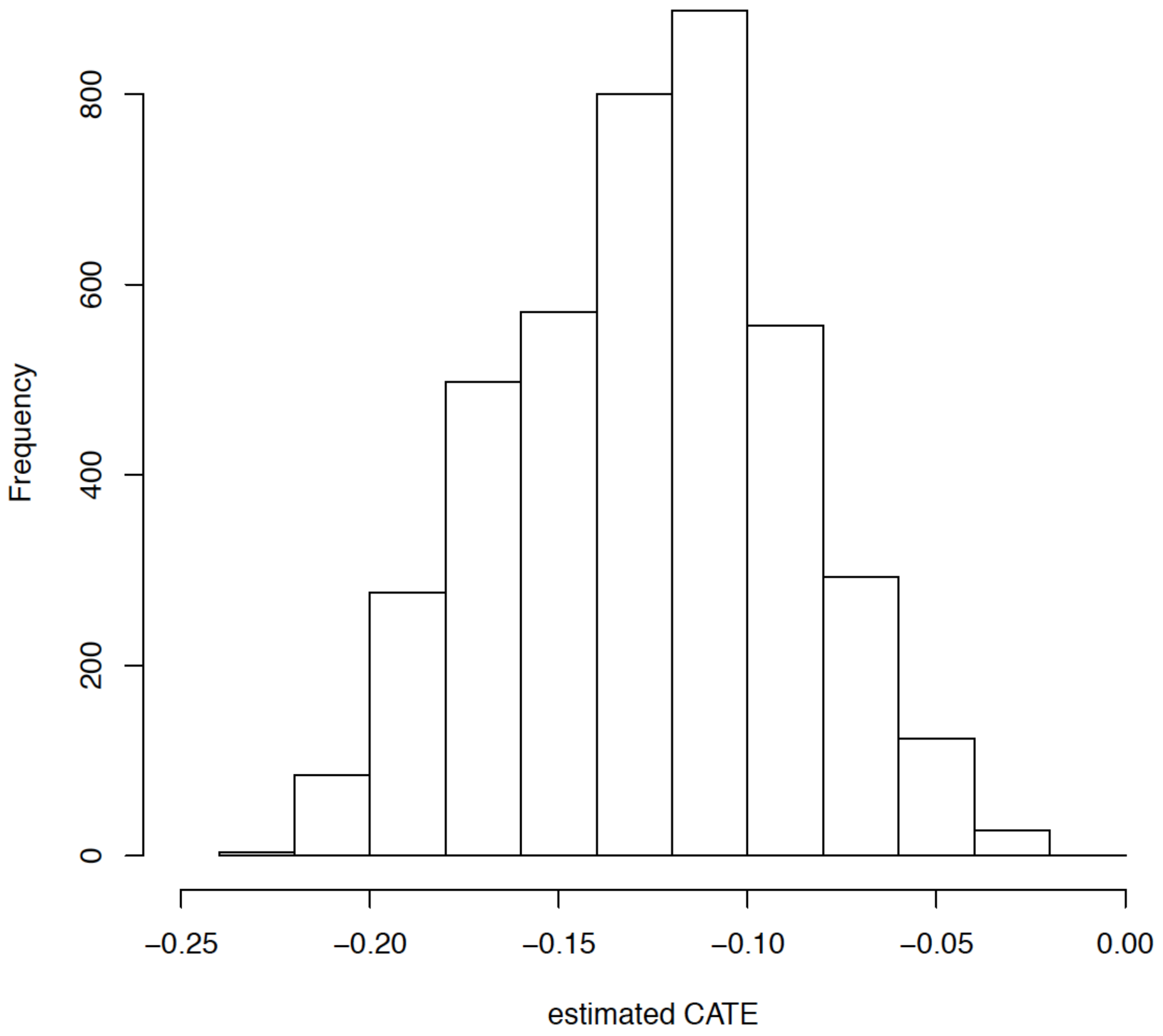

Figure 4. Causal Forest *ITE*s with 95 Percent confidence intervals: ordered by the size of the *ITE*

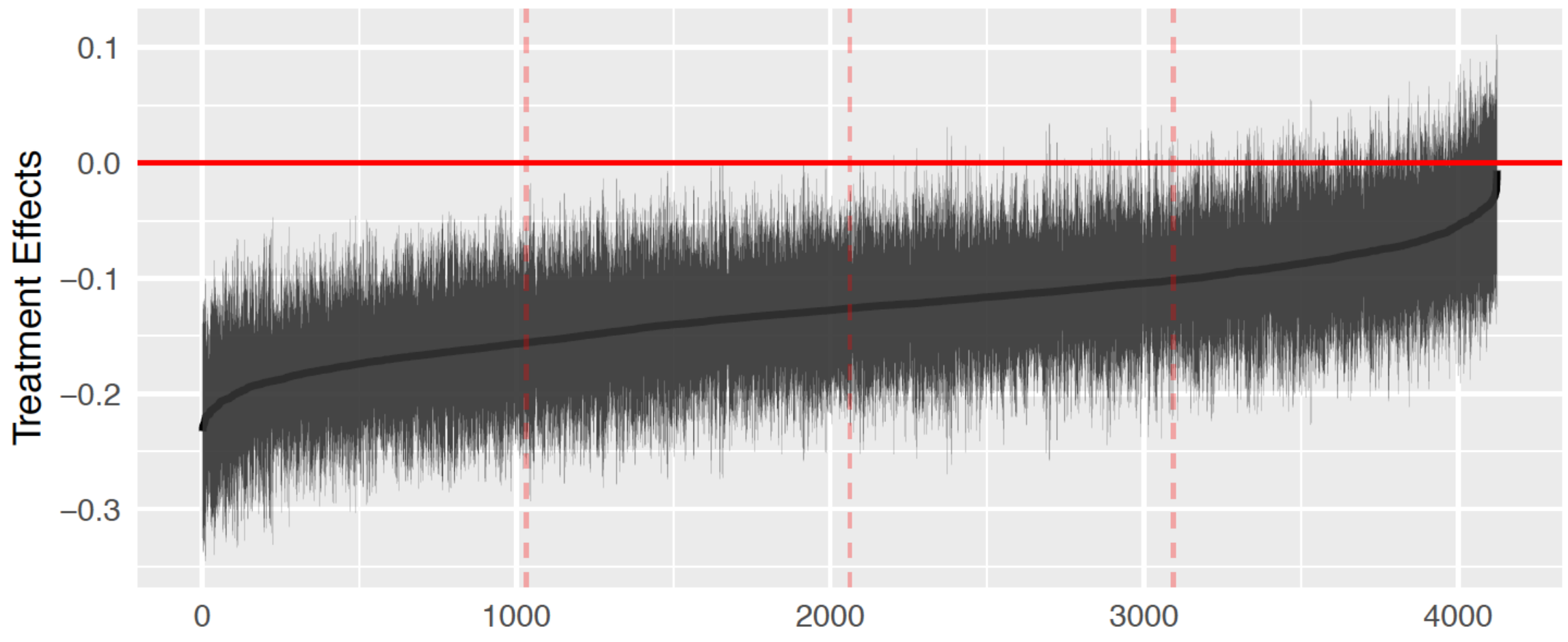

*Notes:* Dashed red lines indicate quartiles of the *ITE*s.

Figure 5. Causal tree for the Effect of college completion on the proportion of time in low wage work

Text in Squares: HTE & sample size (%); Color of Squares: Blue:largest treatment effects & Yellow:smallest treatment effects; Number in Parentheses: Standard Error

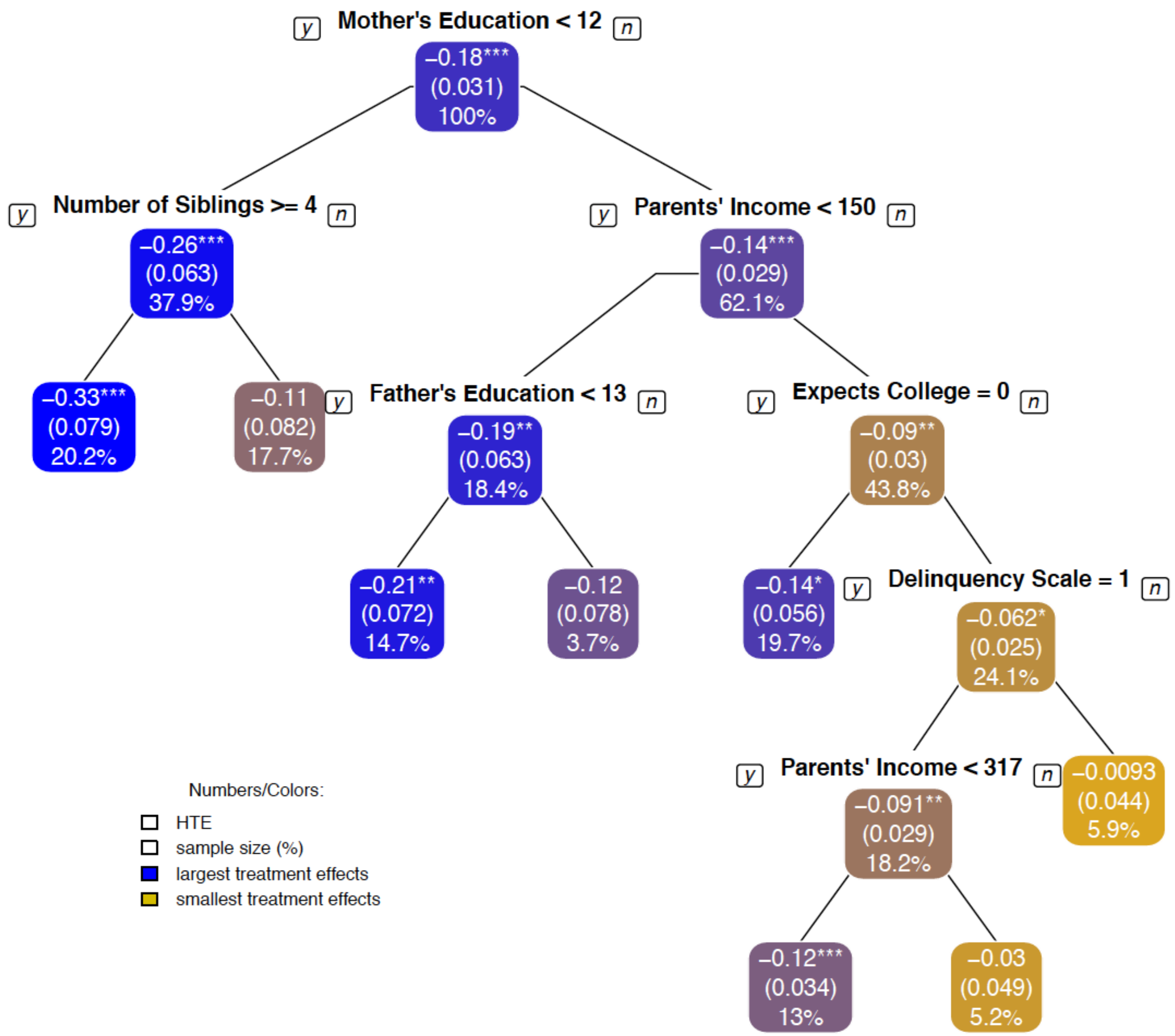

Figure 6. Variable Importance Plot Based on Causal Forest of the Effect of college completion on the proportion of time in low wage work

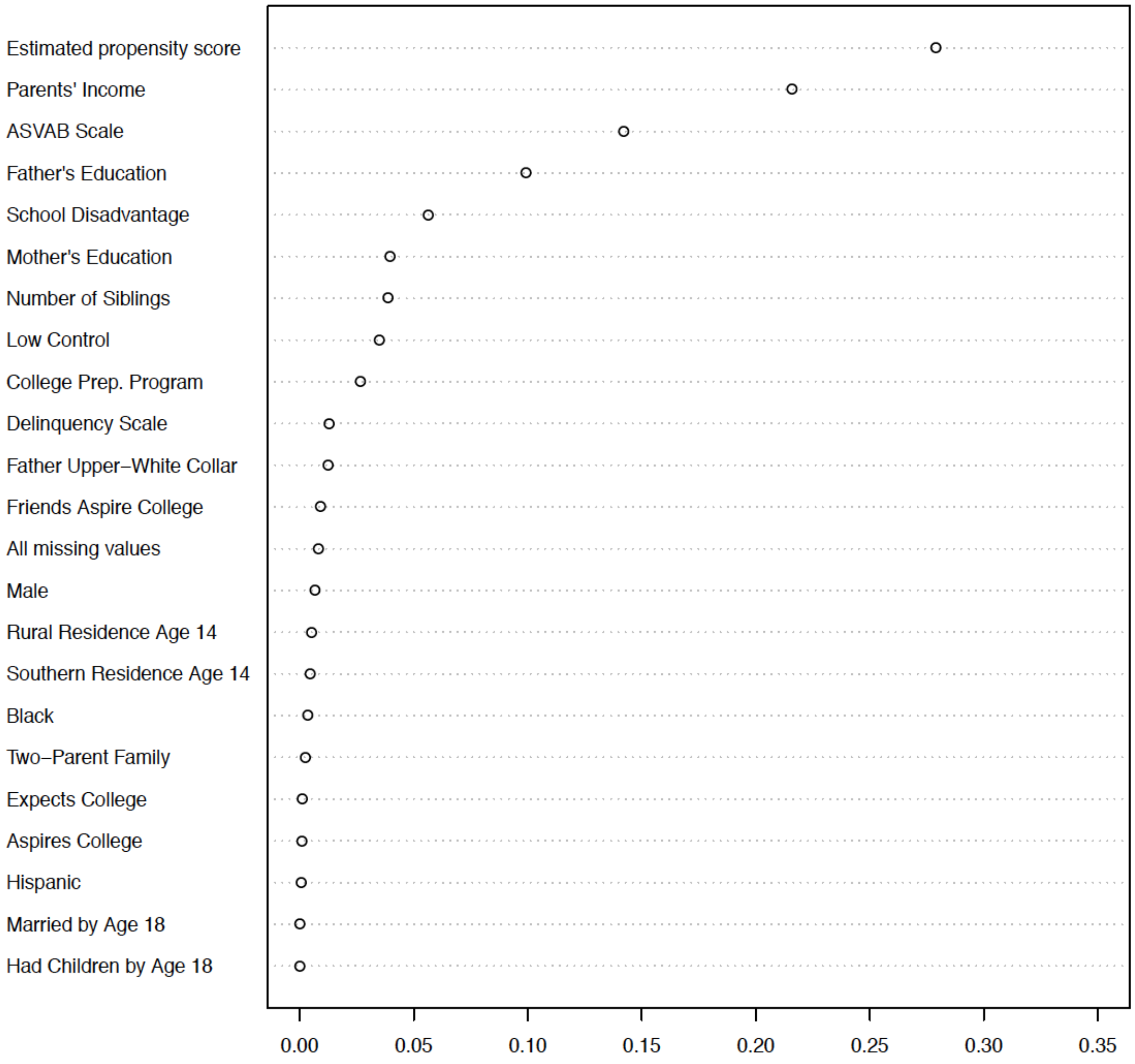

**Figure 7. Depiction of Treatment Effect Heterogeneity by the Propensity of College (*e(x)*) and Unobserved Resistance to College (*u*)**

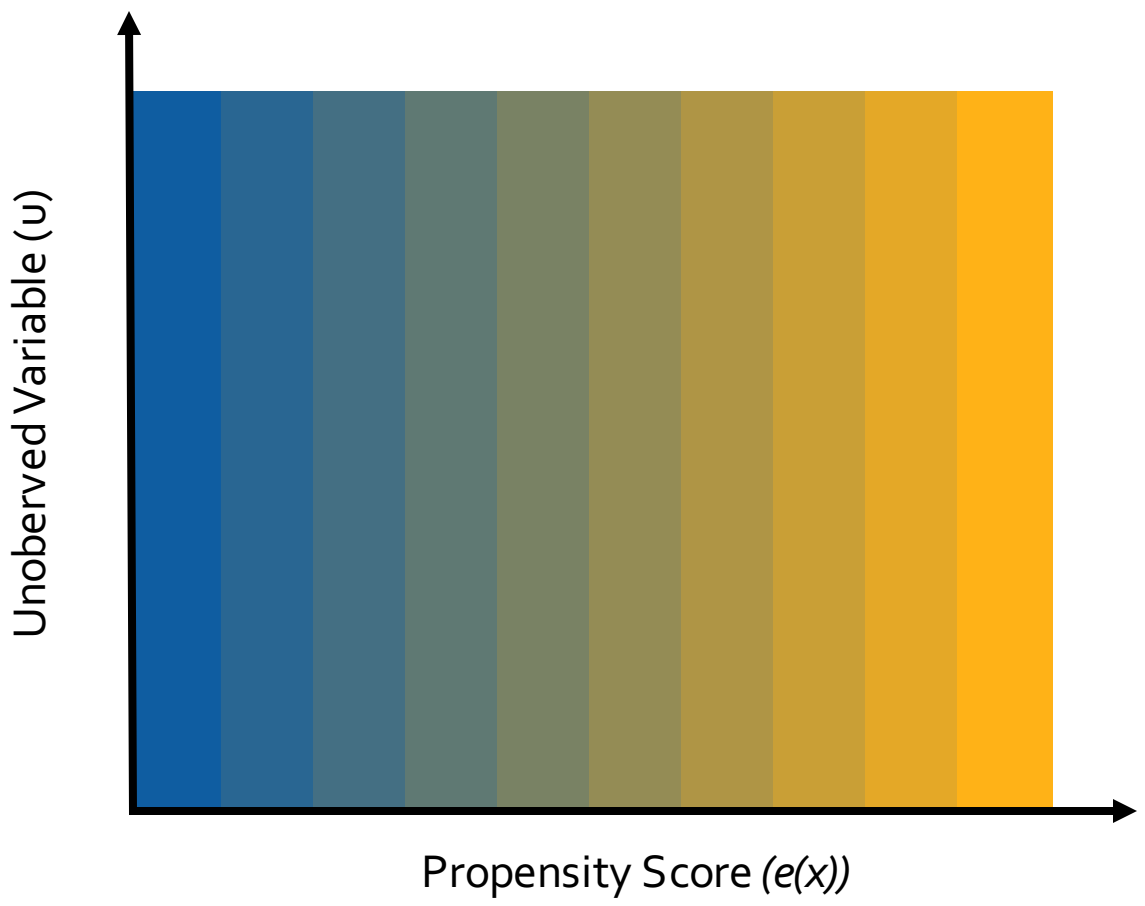

(a) *MTE* for all units, assuming unconfoundedness

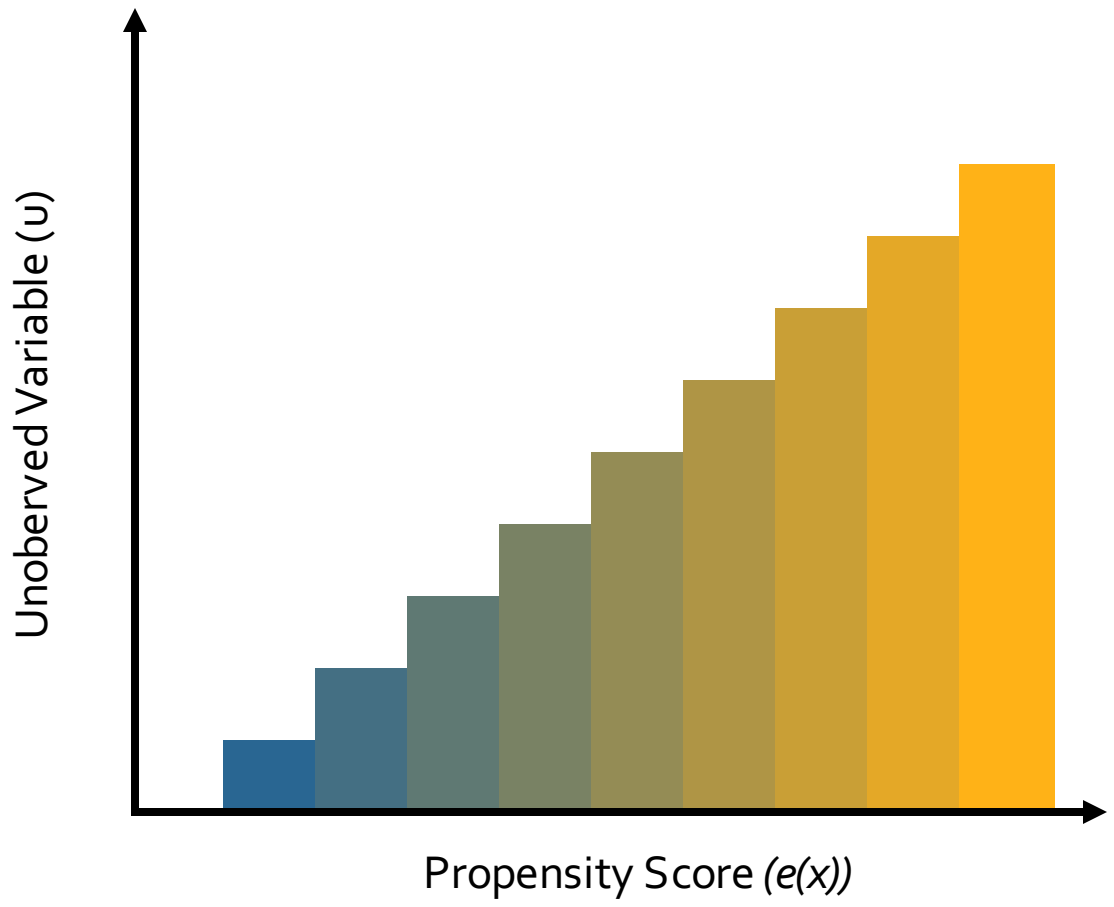

(b) *MTE* for treated units, assuming unconfoundedness

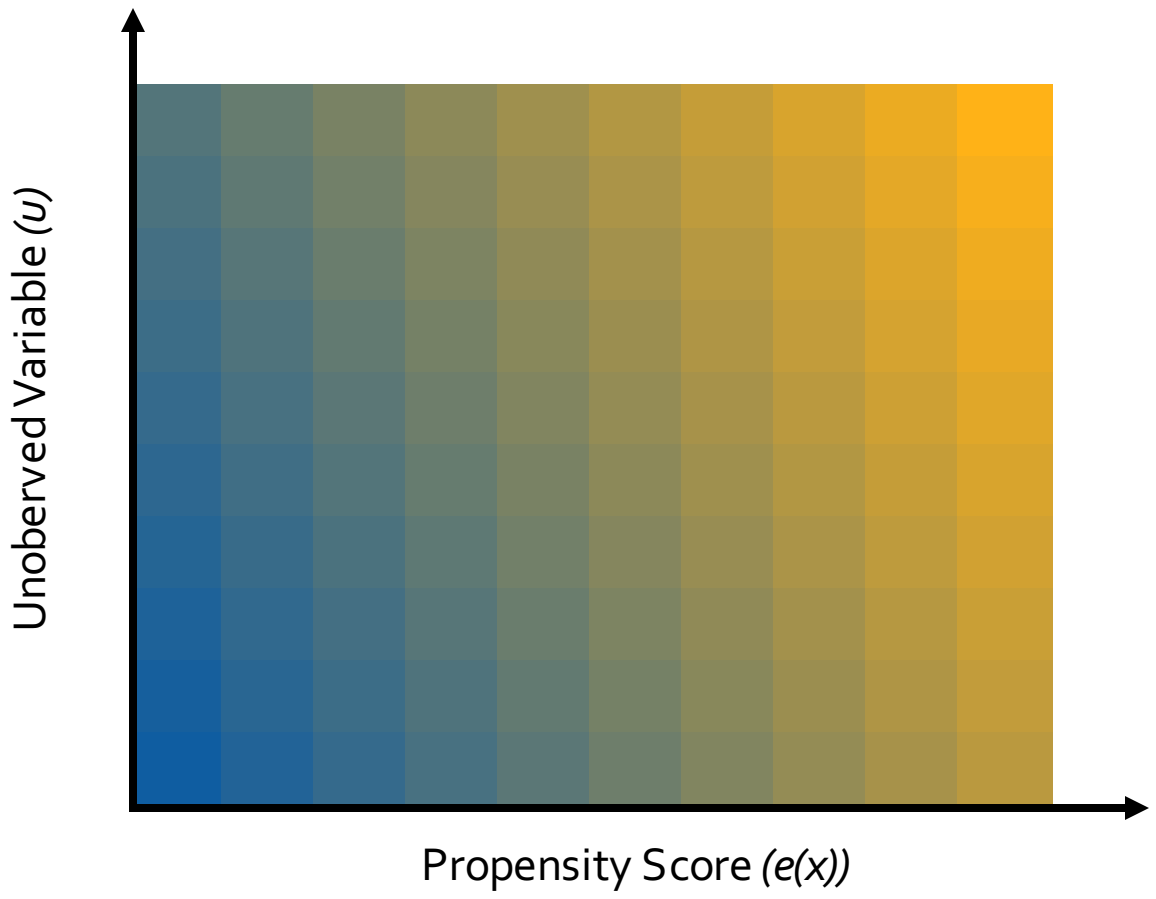

(C) *MTE* for all units

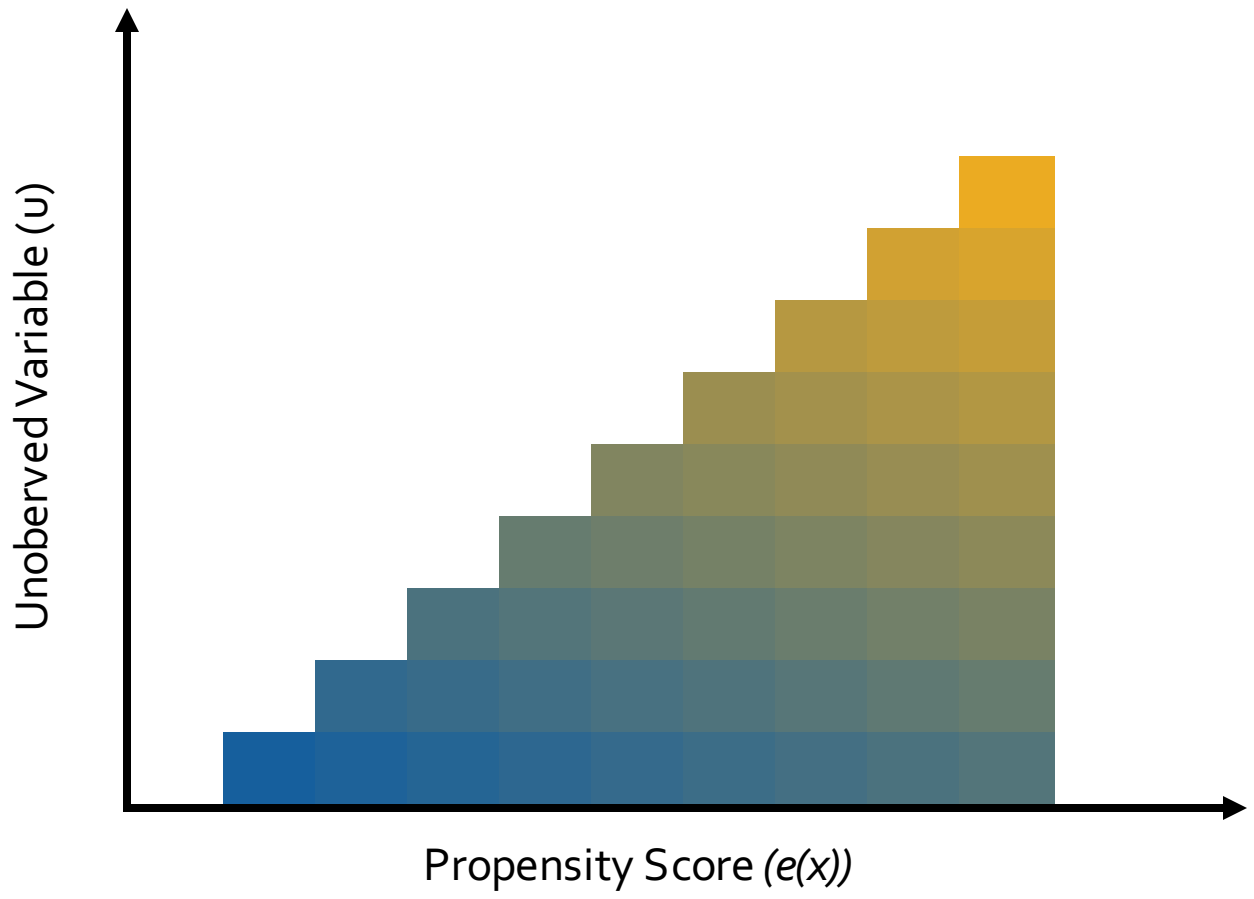

(d) *MTE* for treated units

*Notes:* A darker color indicates a larger treatment effect. Figure includes (a) all units under the unconfoundedness assumption; (b) treated units under the unconfoundedness assumption; (c) all units; and (d) treated units.

**Table 1**

**Descriptive Statistics of Pre-College Characteristics and Wage Outcome**

| | Non College Graduates | | College Graduates | |
|---|---|---|---|---|
| | Mean | (SD) | Mean | (SD) |
| **Sociodemographic Factors** | | | | |
| Male (binary 0/1) | 0.502 | --- | 0.508 | --- |
| Black (binary 0/1) | 0.157 | --- | 0.071 | --- |
| Hispanic (binary 0/1) | 0.056 | --- | 0.028 | --- |
| Southern residence at age 14 (binary 0/1) | 0.320 | --- | 0.292 | --- |
| Rural residence at age 14 (binary 0/1) | 0.245 | --- | 0.191 | --- |
| **Family Background Factors** | | | | |
| Parents' household income ($100s) (continuous 0-75) | 197.686 | (112.375) | 280.746 | (147.107) |
| Fathers' highest education (0-20) | 11.621 | (3.029) | 14.131 | (3.204) |
| Mothers' highest education (0-20) | 11.541 | (2.277) | 13.220 | (2.358) |
| Father upper-white collar occupation (0/1) | 0.193 | --- | 0.488 | --- |
| Two-parent family at age 14 (binary 0/1) | 0.729 | --- | 0.842 | --- |
| Sibship size (continuous 0-19) | 3.209 | (2.188) | 2.554 | (1.650) |
| **Cognitive and Psychosocial Factors** | | | | |
| Cognitive ability ASVAB (continuous -3-3) | -0.054 | (0.639) | 0.586 | (0.544) |
| High school college prepatory program (0/1) | 0.262 | --- | 0.622 | --- |
| Rotter Locus of Control scale (continuous 4-16) | 8.953 | (2.270) | 8.155 | (2.154) |
| Juvenile delinquency activity scale (0-1) | 0.808 | (0.394) | 0.722 | (0.448) |
| Educational expectations (binary 0/1) | 0.344 | --- | 0.822 | --- |
| Educational aspirations (binary 0/1) | 0.474 | --- | 0.879 | --- |
| Friends' educational aspirations (binary 0/1) | 0.390 | --- | 0.732 | --- |
| **School Factors** | | | | |
| School disadvantage scale (0-99) | 20.510 | (17.069) | 13.028 | (12.705) |
| **Family Formation Factors** | | | | |
| Marital status at age 18 (binary 0/1) | 0.027 | --- | 0.003 | --- |
| Had a child by age 18 (binary 0/1) | 0.021 | --- | 0.002 | --- |
| **Wage Outcome** | | | | |
| Prop. of time in low wage work | 0.398 | (0.363) | 0.164 | (0.246) |
| Weighted sample proportion | 0.74 | | 0.26 | |
| *N* | 3,237 | | 848 | |

*Notes:* Data are from the NLS79. Sample is restricted to individuals who were 14-17 years old at the baseline survey in 1979 (n = 5,582) and who had completed at least the 12th grade (n = 4,548), and who have sufficient overlap (4,085). College completion is measured as a four-year degree completed by age 25. All descriptive statistics are weighted by the NLSY sample weight.

**Table 2**

**CATE of College Completion on Wage Outcome**

| | NN Matching | | Causal Forest | |
|---|---|---|---|---|
| **Wage Outcome** | | | | |
| Prop. of time in low wage work | -0.178 | *** | -0.168 | *** |
| | (0.020) | | (0.020) | |

*Notes*: Data are from the NLS79. Sample is restricted to individuals who were 14-17 years old at the baseline survey in 1979 (n = 5,582) and who had completed at least the 12th grade (n = 4,548). College completion is measured as a four-year degree completed by age 25.
* p ≤ 0.05 ** p ≤ 0.01 *** p ≤ 0.001; two-tailed tests;

## Table 3

### *CATE* of College Completion on Wage Outcome: Covariate and Propensity Partioning Matching Results

| | 1 | | 2 | | 3 | |
|---|---|---|---|---|---|---|
| Propensity score strata | -0.220 | *** | -0.126 | *** | -0.062 | * |
| | (0.029) | | (0.025) | | (0.026) | |
| Parental income terciles | -0.276 | *** | -0.192 | *** | -0.070 | |
| | (0.032) | | (0.032) | | (0.053) | |
| Mothers' education | -0.256 | *** | -0.145 | *** | -0.123 | *** |
| | (0.037) | | (0.028) | | (0.024) | |
| Ability terciles | -0.300 | *** | -0.087 | | -0.117 | *** |
| | (0.039) | | (0.059) | | (0.022) | |
| Race | -0.233 | *** | -0.222 | *** | -0.141 | *** |
| | (0.030) | | (0.031) | | (0.027) | |

*Notes*: Data are from the NLS79. Sample is restricted to individuals who were 14-17 years old at the baseline survey in 1979 (n = 5,582) and who had completed at least the 12th grade (n = 4,548). College completion is measured as a four-year degree completed by age 25. Propensity score strata and parental income and ability terciles are 1 for low, 2 for mid, and 3 for high. For mothers' education, 1 indicates less than high school, 2 indicates a high school degree, and 3 indicates some college attendance or more. For race, 1 indicates black, 2 indicates Hispanic, and 3 indicates white (these categories were based on an ordering of the probability of college completion).

* p ≤ 0.05 ** p ≤ 0.01 *** p ≤ 0.001; two-tailed tests;

**Table 4**

**Sesnsitvity Parameters for Propensity and Covariate Partitioning Results**

| | Sensitvity parameters | | 1 | | 2 | | 3 | |
|---|---|---|---|---|---|---|---|---|
| | $\gamma$ | $\lambda$ | CATE | CI | CATE | CI | CATE | CI |
| **Partitions** | | | | | | | | |
| Propensity score | 10% | -10% | -0.210 | (-0.267,-0.153) | -0.116 | (-0.165,-0.067) | -0.052 | (-0.103,-0.001) |
| | 20% | -10% | -0.200 | (-0.257,-0.143) | -0.106 | (-0.155,-0.057) | -0.042 | (-0.093,0.009) |
| | 40% | -10% | -0.180 | (-0.237,-0.123) | -0.086 | (-0.135,-0.037) | -0.022 | (-0.073,0.029) |
| Parental income | 10% | -10% | -0.266 | (-0.329,-0.203) | -0.182 | (-0.245,-0.119) | -0.060 | (-0.164,0.044) |
| | 20% | -10% | -0.256 | (-0.319,-0.193) | -0.172 | (-0.235,-0.109) | -0.050 | (-0.154,0.054) |
| | 40% | -10% | -0.236 | (-0.299,-0.173) | -0.152 | (-0.215,-0.089) | -0.030 | (-0.134,0.074) |
| Mothers' education | 10% | -10% | -0.246 | (-0.319,-0.173) | -0.135 | (-0.190,-0.080) | -0.113 | (-0.160,-0.066) |
| | 20% | -10% | -0.236 | (-0.309,-0.163) | -0.125 | (-0.180,-0.070) | -0.103 | (-0.150,-0.056) |
| | 40% | -10% | -0.216 | (-0.289,-0.143) | -0.105 | (-0.160,-0.050) | -0.083 | (-0.130,-0.036) |
| Ability | 10% | -10% | -0.290 | (-0.366,-0.214) | -0.077 | (-0.193,0.039) | -0.107 | (-0.150,-0.064) |
| | 20% | -10% | -0.280 | (-0.356,-0.204) | -0.067 | (-0.183,0.049) | -0.097 | (-0.140,-0.054) |
| | 40% | -10% | -0.260 | (-0.336,-0.184) | -0.047 | (-0.163,0.069) | -0.077 | (-0.120,-0.034) |
| Race | 10% | -10% | -0.223 | (-0.282,-0.164) | -0.212 | (-0.273,-0.151) | -0.131 | (-0.184,-0.078) |
| | 20% | -10% | -0.213 | (-0.272,-0.154) | -0.202 | (-0.263,-0.141) | -0.121 | (-0.174,-0.068) |
| | 40% | -10% | -0.193 | (-0.252,-0.134) | -0.182 | (-0.243,-0.121) | -0.101 | (-0.154,-0.048) |

*Notes:* Data are from the NLS79. Sample is restricted to individuals who were 14-17 years old at the baseline survey in 1979 (n = 5,582) and who had completed at least the 12th grade (n = 4,548). College completion is measured as a four-year degree completed by age 25. Propensity score strata and parental income and ability terciles are 1 for low, 2 for mid, and 3 for high. For mothers' education, 1 indicates less than high school, 2 indicates a high school degree, and 3 indicates some college attendance or more. For race, 1 indicates black, 2 indicates Hispanic, and 3 indicates white (these categories were based on an ordering of the probability of college completion).

## Table 5
## Causal Forest *ITE* Quartiles

| | | Quartile 1 | Quartle 2 | Qaurtle 3 | Quartile 4 |
|---|---|---|---|---|---|
| Predicted effect of college on low wage job | | -0.25 | -0.16 | -0.13 | -0.10 |
| Mean prop. of time in a low wage job | | 0.51 | 0.38 | 0.31 | 0.25 |
| Treated group prop. of time in a low wage job | | 0.21 | 0.24 | 0.20 | 0.20 |
| Var. prop. of time in a low wage job | | 0.14 | 0.13 | 0.11 | 0.10 |
| Var. treated group of time in a low wage job | | 0.07 | 0.10 | 0.08 | 0.07 |
| Max prop. of time in a low wage job | | 1.00 | 1.00 | 1.00 | 1.00 |
| Min prop. of time in a low wage job | | 0.00 | 0.00 | 0.00 | 0.00 |
| Propensity Score | | | | | |
| | Low | 24.3% | 20.7% | 14.7% | 6.3% |
| | Mid | 0.5% | 3.7% | 8.5% | 5.4% |
| | High | 0.2% | 0.6% | 1.8% | 13.3% |
| Parental income | | | | | |
| | Low | 15.8% | 10.2% | 6.4% | 0.9% |
| | Mid | 8.3% | 10.8% | 10.3% | 4.6% |
| | High | 0.9% | 4.0% | 8.3% | 19.5% |
| Mothers' education | | | | | |
| | < HS | 16.2% | 11.6% | 7.4% | 2.7% |
| | HS | 7.4% | 9.8% | 13.1% | 12.9% |
| | College | 1.4% | 3.6% | 4.5% | 9.4% |
| Ability | | | | | |
| | Low | 17.6% | 8.6% | 4.2% | 3.0% |
| | Mid | 6.2% | 10.0% | 9.6% | 7.6% |
| | High | 1.2% | 6.4% | 11.2% | 14.5% |
| Race | | | | | |
| | Black | 10.8% | 8.2% | 6.9% | 1.9% |
| | Hispanic | 4.4% | 4.8% | 3.2% | 1.8% |
| | White | 9.8% | 11.9% | 14.9% | 21.3% |

*Notes:* Data are from the NLS79. Sample is restricted to individuals who were 14-17 years old at the baseline survey in 1979 (n = 5,582) and who had completed at least the 12th grade (n = 4,548). College completion is measured as a four-year degree completed by age 25.

## Table 6

### Propensity Scores and Balance Metrics by College Completion for Recursive Partitions

| | Mean propensity score | | Propensity balance metric | |
|---|---|---|---|---|
| | Non-college graduate | College graduate | Raw sample | Matched sample |
| L1: Mothers' education < 12 | 0.074 | 0.309 | 1.592 | 0.078 |
| L2: Number of silbings >= 4 | 0.061 | 0.223 | 1.424 | 0.105 |
| L3: Number of siblings < 4 | 0.137 | 0.358 | 1.641 | 0.220 |
| L4: Mother's education >= 12 | 0.180 | 0.527 | 1.575 | 0.034 |
| L5: Parents income < 150 | 0.123 | 0.371 | 1.450 | 0.100 |
| L6: Fathers' education < 13 | 0.104 | 0.319 | 1.399 | 0.124 |
| L7: Fathers' education >= 13 | 0.220 | 0.449 | 1.159 | 0.220 |
| L8: Parents' income >= 150 | 0.209 | 0.562 | 1.544 | 0.045 |
| L9: Expects college = 0 | 0.090 | 0.251 | 1.325 | 0.135 |
| L10: Expects college = 1 | 0.378 | 0.623 | 1.112 | 0.052 |
| L11: Delinquency scale = 1 | 0.364 | 0.616 | 1.139 | 0.060 |
| L12: Parents' income < 317 | 0.328 | 0.545 | 1.071 | 0.085 |
| L13: Parents' income >= 317 | 0.550 | 0.712 | 0.779 | 0.056 |
| L14: Delinquency = 0 | 0.434 | 0.641 | 0.963 | 0.084 |

*Notes*: Data are from the NLS79. Sample is restricted to individuals who were 14-17 years old at the baseline survey in 1979 (n = 5,582) and who had completed at least the 12th grade (n = 4,548). College completion is measured as a four-year degree completed by age 25. Balance metrics are based on the linearized propensity scores.
* $p \leq 0.05$ ** $p \leq 0.01$ *** $p \leq 0.001$; two-tailed tests;

## Table 7

### Sesnsitvity Parameters for Recursive Partitioning Results

| | Sensitvity parameters | | | |
|---|---|---|---|---|
| | $\gamma$ | $\lambda$ | CATE | CI |
| **Partitions** | | | | |
| L1: Mothers' education < 12 | 10% | -10% | -0.247 | (-0.369,-0.124) |
| | 20% | -10% | -0.237 | (-0.359,-0.114) |
| | 40% | -10% | -0.217 | (-0.339,-0.094) |
| L2: Number of silbings >= 4 | 10% | -10% | -0.319 | (-0.474,-0.164) |
| | 20% | -10% | -0.309 | (-0.464,-0.154) |
| | 40% | -10% | -0.289 | (-0.444,-0.134) |
| L3: Number of siblings < 4 | 10% | -10% | -0.097 | (-0.257,0.064) |
| | 20% | -10% | -0.087 | (-0.247,0.074) |
| | 40% | -10% | -0.067 | (-0.227,0.094) |
| L4: Mother's education >= 12 | 10% | -10% | -0.126 | (-0.184,-0.069) |
| | 20% | -10% | -0.116 | (-0.174,-0.059) |
| | 40% | -10% | -0.096 | (-0.154,-0.039) |
| L5: Parents income < 150 | 10% | -10% | -0.185 | (-0.309,-0.060) |
| | 20% | -10% | -0.175 | (-0.299,-0.050) |
| | 40% | -10% | -0.155 | (-0.279,-0.030) |
| L6: Fathers' education < 13 | 10% | -10% | -0.203 | (-0.343,-0.062) |
| | 20% | -10% | -0.193 | (-0.333,-0.052) |
| | 40% | -10% | -0.173 | (-0.313,-0.032) |
| L7: Fathers' education >= 13 | 10% | -10% | -0.110 | (-0.263,0.043) |
| | 20% | -10% | -0.100 | (-0.253,0.053) |
| | 40% | -10% | -0.080 | (-0.233,0.073) |
| L8: Parents' income >= 150 | 10% | -10% | -0.080 | (-0.139,-0.022) |
| | 20% | -10% | -0.070 | (-0.129,-0.012) |
| | 40% | -10% | -0.050 | (-0.109,0.008) |
| L9: Expects college = 0 | 10% | -10% | -0.127 | (-0.155,-0.098) |
| | 20% | -10% | -0.117 | (-0.145,-0.088) |
| | 40% | -10% | -0.097 | (-0.125,-0.068) |
| L10: Expects college = 1 | 10% | -10% | -0.053 | (-0.102,-0.003) |
| | 20% | -10% | -0.043 | (-0.092,0.007) |
| | 40% | -10% | -0.023 | (-0.072,0.027) |
| L11: Delinquency scale = 1 | 10% | -10% | -0.081 | (-0.137,-0.024) |
| | 20% | -10% | -0.071 | (-0.127,-0.014) |
| | 40% | -10% | -0.051 | (-0.107,0.006) |
| L12: Parents' income < 317 | 10% | -10% | -0.105 | (-0.173,-0.038) |
| | 20% | -10% | -0.095 | (-0.163,-0.028) |
| | 40% | -10% | -0.075 | (-0.143,-0.008) |
| L13: Parents' income >= 317 | 10% | -10% | -0.020 | (-0.116,0.076) |
| | 20% | -10% | -0.010 | (-0.106,0.086) |
| | 40% | -10% | 0.010 | (-0.086,0.106) |
| L14: Delinquency scale = 0 | 10% | -10% | 0.001 | (-0.086,0.087) |
| | 20% | -10% | 0.011 | (-0.076,0.097) |
| | 40% | -10% | 0.031 | (-0.056,0.117) |

*Notes:* Data are from the NLS79. Sample is restricted to individuals who were 14-17 years old at the baseline survey in 1979 (n = 5,582) and who had completed at least the 12th grade (n = 4,548). College completion is measured as a four-year degree completed by age 25.

## Appendix Table A

### Logit Regression Estimates Predicting College Completion

| | β | (SE) | |
|---|---|---|---|
| **Sociodemographic factors** | | | |
| Male (binary 0/1) | -0.884 | (0.228) | *** |
| Black (binary 0/1) | -0.300 | (0.188) | |
| Hispanic (binary 0/1) | -0.725 | (0.196) | *** |
| Southern residence at age 14 (binary 0/1) | -0.287 | (0.177) | |
| Rural residence at age 14 (binary 0/1) | 0.236 | (0.167) | |
| **Family background factors** | | | |
| Parents' household income ($100s) (continuous 0-75) | -0.004 | (0.001) | ** |
| Fathers' highest education (0-20) | 0.027 | (0.021) | |
| Mothers' highest education (0-20) | -0.088 | (0.094) | |
| Father upper-white collar occupation (0/1) | 0.831 | (0.180) | *** |
| Two-parent family at age 14 (binary 0/1) | 0.215 | (0.124) | |
| Sibship size (continuous 0-19) | -0.143 | (0.033) | *** |
| **Cognitive and psychosocial factors** | | | |
| Cognitive ability ASVAB (continuous -3-3) | 1.378 | (0.124) | *** |
| High school college prepatory program (0/1) | 0.918 | (0.164) | *** |
| Rotter Locus of Control scale (continuous 4-16) | -0.034 | (0.022) | |
| Juvenile delinquency activity scale (0-1) | -0.886 | (0.220) | *** |
| Educational expectations (binary 0/1) | 1.380 | (0.592) | ** |
| Educational aspirations (binary 0/1) | 2.578 | (0.695) | *** |
| Friends' educational aspirations (binary 0/1) | 0.286 | (0.109) | ** |
| **School factors** | | | |
| School disadvantage scale (0-99) | -0.006 | (0.003) | |
| **Family formation factors** | | | |
| Marital status at age 18 (binary 0/1) | -2.501 | (0.736) | ** |
| Had a child by age 18 (binary 0/1) | -1.896 | (0.629) | ** |
| **Higher order terms** | | | |
| Mothers' education x mothers' education | 0.014 | (0.003) | *** |
| Sibship size x southern residence | 0.827 | (0.048) | *** |
| Male x educational aspirations | 0.827 | (0.252) | ** |
| Parents' income x delinquent activity | 0.003 | (0.001) | ** |
| Mothers' education X educational aspirations | -0.195 | (0.056) | ** |
| Parental income x educational expectations | 0.003 | (0.001) | ** |
| Hispanic x southern residence | -0.819 | (0.345) | * |
| Ability x high school program | -0.353 | (0.168) | * |
| Educational expectations x aspirations | -1.302 | (0.588) | * |
| Father upper-white x high school program | -0.587 | (0.239) | * |
| Black x high school program | -0.582 | (0.239) | * |
| Rural residence x high school program | -0.511 | (0.239) | * |
| **Missing indicator** | | | |
| Missing values imputed | -0.040 | (0.108) | |
| *N* | 4548 | | |
| *Log Likelihood* | 1667.6 | | |
| $P > \chi^2$ | 0.000 | | |

*Notes*: Data are from the NLS79. Sample is restricted to individuals who were 14-17 years old at the baseline survey in 1979 (n = 5,582) and who had completed at least the 12th grade (n = 4,548). College completion is measured as a four-year degree completed by age 25.
* $p \leq 0.05$ ** $p \leq 0.01$ *** $p \leq 0.001$; two-tailed tests;

## Appendix Table B

### Tests of Significance between Propensity and Covariate Partitioning Results

| | | Propensity score | | | Parental income | | | Mothers' education | | | Ability | | | Race | | |
|---|---|---|---|---|---|---|---|---|---|---|---|---|---|---|---|---|
| | | Low | Mid | High | Low | Mid | High | < HS | HS | College | Low | Mid | High | Black | Hisp. | White |
| Propensity score | Low | | | | | | | | | | | | | | | |
| | Mid | -76.99 | | | | | | | | | | | | | | |
| | High | -126.23 | -51.08 | | | | | | | | | | | | | |
| Parental income | Low | 34.07 | 88.36 | 128.19 | | | | | | | | | | | | |
| | Mid | -32.49 | 40.20 | 88.83 | -55.00 | | | | | | | | | | | |
| | High | -120.32 | -40.12 | 12.98 | -121.98 | -80.13 | | | | | | | | | | |
| Mothers' education | < HS | 14.18 | 72.75 | 114.81 | -17.01 | 37.80 | 107.89 | | | | | | | | | |
| | HS | -70.86 | 12.56 | 65.79 | -82.95 | -30.72 | 55.29 | -66.29 | | | | | | | | |
| | College | -83.60 | -4.04 | 48.15 | -93.11 | -45.10 | 36.87 | -77.51 | -17.19 | | | | | | | |
| Ability | Low | 49.18 | 97.05 | 133.39 | 16.19 | 67.15 | 127.46 | 32.09 | 92.16 | 101.30 | | | | | | |
| | Mid | -89.91 | -12.50 | 38.99 | -98.37 | -52.39 | 27.41 | -83.29 | -25.81 | -8.73 | -106.09 | | | | | |
| | High | -125.06 | -15.90 | 46.11 | -116.11 | -66.55 | 33.18 | -99.80 | -34.03 | -11.34 | -120.90 | -0.36 | | | | |
| Race | Black | 0.76 | 55.72 | 96.08 | -25.54 | 23.46 | 88.44 | -10.04 | 48.54 | 59.76 | -39.12 | 65.62 | 76.07 | | | |
| | Hispanic | 2.97 | 43.76 | 75.76 | -17.55 | 19.62 | 68.72 | -5.45 | 37.60 | 46.54 | -29.11 | 51.35 | 55.57 | 2.13 | | |
| | White | -90.00 | 8.64 | 66.77 | -94.01 | -39.87 | 55.79 | -76.76 | -5.95 | 13.74 | -101.71 | 23.34 | 33.54 | -56.06 | -41.83 | |

Notes: Data are from the NLS79. Sample is restricted to individuals who were 14-17 years old at the baseline survey in 1979 (n = 5,582) and who had completed at least the 12th grade (n = 4,548). College completion is measured as a four-year degree completed by age 25. Cells indicate t-test values for tests of difference between each of the pairs of subgroup effects.

* $p \leq 0.05$ ** $p \leq 0.01$ *** $p \leq 0.001$; two-tailed tests;

**Appendix Table C**

***CATE* of College Completion on Wage Outcome: Recursive Partitioning Matching Results**

| | | | *N* |
|---|---|---|---|
| L1: Mothers' education < 12 | -0.257 | *** | 1547 |
| | (0.063) | | |
| L2: Number of silbings >= 4 | -0.329 | *** | 824 |
| | (0.079) | | |
| L3: Number of siblings < 4 | -0.107 | | 723 |
| | (0.082) | | |
| L4: Mother's education >= 12 | -0.136 | *** | 2538 |
| | (0.029) | | |
| L5: Parents income < 150 | -0.195 | ** | 750 |
| | (0.063) | | |
| L6: Fathers' education < 13 | -0.213 | ** | 600 |
| | (0.072) | | |
| L7: Fathers' education >= 13 | -0.120 | | 150 |
| | (0.078) | | |
| L8: Parents' income >= 150 | -0.090 | ** | 1788 |
| | (0.030) | | |
| L9: Expects college = 0 | -0.137 | * | 805 |
| | (0.014) | | |
| L10: Expects college = 1 | -0.063 | * | 983 |
| | (0.025) | | |
| L11: Delinquency scale = 1 | -0.091 | ** | 744 |
| | (0.029) | | |
| L12: Parents' income < 317 | -0.115 | *** | 532 |
| | (0.034) | | |
| L13: Parents' income >= 317 | -0.030 | | 212 |
| | (0.049) | | |
| L14: Delinquency = 0 | -0.009 | | 239 |
| | (0.044) | | |

*Notes*: Data are from the NLS79. Sample is restricted to individuals who were 14-17 years old at the baseline survey in 1979 (n = 5,582) and who had completed at least the 12th grade (n = 4,548). College completion is measured as a four-year degree completed by age 25.
* p ≤ 0.05 ** p ≤ 0.01 *** p ≤ 0.001; two-tailed tests;

## Appendix Table D
## Tests of Significance between Recursive Partitioning Results

| Leaf legend | Leaf | 1 | 2 | 3 | 4 | 5 | 6 | 7 | 8 | 9 | 10 | 11 | 12 | 13 | 14 |
|---|---|---|---|---|---|---|---|---|---|---|---|---|---|---|---|
| L1: Mothers' education < 12 | 1 | | | | | | | | | | | | | | |
| L2: Number of silbings >= 4 | 2 | 22.73 | | | | | | | | | | | | | |
| L3: Number of siblings < 4 | 3 | -43.65 | -54.12 | | | | | | | | | | | | |
| L4: Mother's education >= 12 | 4 | -71.11 | -68.41 | 9.51 | | | | | | | | | | | |
| L5: Parents income < 150 | 5 | -22.10 | -37.34 | 22.98 | 24.47 | | | | | | | | | | |
| L6: Fathers' education < 13 | 6 | -13.29 | -29.03 | 25.08 | 25.64 | 4.80 | | | | | | | | | |
| L7: Fathers' education >= 13 | 7 | -20.79 | -30.10 | 1.91 | -2.50 | -10.98 | -13.17 | | | | | | | | |
| L8: Parents' income >= 150 | 8 | -95.60 | -83.86 | -5.25 | -50.23 | -43.10 | -40.69 | -4.67 | | | | | | | |
| L9: Expects college = 0 | 9 | -71.83 | -68.58 | 9.72 | 0.65 | -24.43 | -25.58 | 2.58 | 53.43 | | | | | | |
| L10: Expects college = 1 | 10 | -108.74 | -92.71 | -14.02 | -73.90 | -53.86 | -49.52 | -8.99 | -25.89 | -77.62 | | | | | |
| L11: Delinquency scale = 1 | 11 | -87.00 | -80.73 | -4.97 | -37.85 | -40.86 | -39.25 | -4.57 | 0.32 | -39.38 | 21.24 | | | | |
| L12: Parents' income < 317 | 12 | -64.83 | -68.17 | 2.54 | -13.06 | -28.80 | -29.65 | -0.75 | 15.16 | -13.58 | 31.11 | 13.48 | | | |
| L13: Parents' income >= 317 | 13 | -60.77 | -68.62 | -16.86 | -31.01 | -40.22 | -40.90 | -12.50 | -17.47 | -31.26 | -9.34 | -17.15 | -23.11 | | |
| L14: Delinquency = 0 | 14 | -75.74 | -80.58 | -23.34 | -43.59 | -50.44 | -49.79 | -15.89 | -27.57 | -43.97 | -17.94 | -26.77 | -32.93 | -4.71 | |

*Notes:* Data are from the NLS79. Sample is restricted to individuals who were 14-17 years old at the baseline survey in 1979 (n = 5,582) and who had completed at least the 12th grade (n = 4,548). College completion is measured as a four-year degree completed by age 25. Cells indicate t-test values for tests of difference between each of the pairs of leaves represented by the leaf number.

* p ≤ 0.05 ** p ≤ 0.01 *** p ≤ 0.001; two-tailed tests;